\documentclass[aps,reprint,prd,showpacs,amsmath,amssymb,nofootinbib]{revtex4-1}
\usepackage{graphicx}
\usepackage{color}
\usepackage{times}
\usepackage{inputenc}
\usepackage{bm}
\usepackage{multirow}
\usepackage{float}
\usepackage{url}
\usepackage{natbib}
\usepackage{tensor}
\usepackage[colorlinks=true,citecolor=blue,urlcolor=blue,linkcolor=red]{hyperref}
\usepackage[normalem]{ulem}
\usepackage{lineno}
\usepackage{ulem}

\def\msun{\,M_{\odot}}
\def\fm3{\;\text{fm}^{-3}}
\def\mev{\;\text{MeV}}

\newcommand{\green}[1]{{\color[rgb]{0.0,0.7,0.0}#1}}

\begin{document}
\title{Hybrid stars with large quark cores within the parity doublet model and modified NJL model }
\author{Wen-Li Yuan$^{1}$}
\email{wlyuan@pku.edu.cn} 
\author{Bikai Gao$^{2,3}$}
\email{gaobikai@hken.phys.nagoya-u.ac.jp}
\author{Yan Yan$^{4}$}
\author{Renxin Xu$^{1}$}
\email{r.x.xu@pku.edu.cn}

\affiliation{$^1$School of Physics and State Key Laboratory of Nuclear Physics and Technology, Peking University, Beijing 100871, China;\\ 
$^2$Department of Physics, Nagoya University, Nagoya 464-8602, Japa6;\\
$^3$Research Center for Nuclear Physics (RCNP), Osaka University, Osaka 567-0047, Japan;\\
$^4$School of Microelectronics and Control Engineering, Changzhou University, Changzhou 213164, China
}

\date{\today} 

\begin{abstract}
Using the parity doublet model (PDM) for hadronic matter and a modified Nambu-Jona-Lasinio (NJL) model for quark matter, we investigate the potential existence of two- and three-flavor quark matter in neutron star cores. Both models respect chiral symmetry, and a sharp first-order phase transition is implemented via Maxwell construction. We find stable neutron stars with quark cores within a specific parameter space that satisfies current astronomical observations. Typical neutron stars with masses around $1.4\msun$ may possess deconfined quark matter in their centers. The hybrid star scenario with a two-flavor quark core offers enough parameter space to allow the neutron stars with large quark cores exceeding $\sim 1\msun$, and allow the early deconfinement position before $2~\rho_0$, where $\rho_0$ is the nuclear saturation density. The observations of gravitational wave event GW170817 suggest a relatively large chiral invariant mass $m_0=600~\mev$ in the PDM for scenarios involving three-flavor quark matter cores. The maximum mass of the hybrid star with a quark core is found to be approximately $2.2~M_\odot$ for both two- or three-flavor quark matter in their centers. 

\end{abstract}
\maketitle 

\section{Introduction}
 The study of quantum chromodynamics (QCD) phase transitions is a hot topic in hadron physics, driving extensive theoretical and experimental investigations. A central goal of ultrarelativistic heavy-ion collision experiments is to explore the nature of the hadron-quark phase transition. As temperature and density increase, strongly interacting matter is expected to undergo a transition from the hadronic phase to quark-gluon plasma~\cite{2005PhR...407..205B,2008PhRvD..77k4028F,2011RPPh...74a4001F,2017EPJWC.14104001L}. However, key questions including the existence and precise location of the critical endpoint in the QCD phase diagram remain unresolved~\cite{2002LNP...583..209K,2003ARNPS..53..163L, 2007PhRvD..76g4023S,2017ApJ...836...89K}. It is even unclear whether there are real phase transitions or only crossover transitions~\cite{1985ARNPS..35..245S,1992PhRvL..69..737C,2000PhLB..486..239S,2007JPhG...34.2655H,2011PhRvD..83i4033S,2012JPhG...39a3101F,2013PrPNP..72...99F,2016PhRvD..94g6009G,2019EPJC...79..399X,2023Symm...15..541H}. Such transitions appeal great interest not only in the context of heavy-ion physics but also in the study of neutron stars, which provide a unique extreme astrophysical environment for exploring cold matter at supranuclear densities~\cite{2002PhRvC..66b5802B,2005ApJ...629..969A,2007Natur.445E...7A,2013PhRvD..88h3013A,2013ApJ...773...11H,2023PhRvC.108b5803Z,2020NatPh..16..907A,2023NatCo..14.8451A,2024PhRvD.109f3035C,2020ApJ...904..103M,2022ApJ...935...88H,2023ApJ...944..206L,2024PhRvC.110d5802G}. Theoretically, both strange and nonstrange quark matter are expected to reside in neutron stars~\cite{2005PhR...407..205B,2008PhRvD..77k4028F,2011RPPh...74a4001F,2017EPJWC.14104001L}, where the behavior of dense QCD matter can be probed, potentially shedding light on the nature of phase transitions. 

In recent years, the advancement of the multimessenger astronomy era has provided deeper insights into neutron star physics. For comprehensive reviews of this topic, see, e.g., Refs.~\cite{2008RvMP...80.1455A,2018RPPh...81e6902B,2024LRR....27....3K}. Observations of neutron star mass and radius encode unique information on the equation of state (EOS) at supranuclear densities, offering critical insights into the properties of QCD matter~\cite{2015PhRvD..91d5003K,2020arXiv201110940K,2021PhRvD.104g4005K,2021ApJ...923..250J,2023PhRvD.108k4028G,2024PhRvD.109l3005M,2024arXiv241114938Y,2025arXiv250100115A,2025arXiv250313626C}. The discovery of several massive pulsars with masses close to two solar masses, such as $M = 1.908 \pm 0.016\msun$ for PSR J1614-2230~\cite{2010Natur.467.1081D}, $M = 2.01 \pm 0.04\msun$ for PSR J0348+0432~\cite{2013Sci...340..448A}, and $M = 2.14^{+0.10}_{-0.09}\msun$ for PSR J0740+6620~\cite{2020NatAs...4...72C}, has sparked considerable interests and discussions about the possible presence of quark matter in the cores of neutron stars~\cite{2015PhRvC..91c5803L,2020PhRvD.102h3030F,2020PhRvD.101l3030F,2022JPhG...49d5201L,2023SciBu..68..913H,2023PhRvD.107i4032L,2024PhRvD.109d3052F}. The presence of a phase transition to quark matter, especially a first-order phase transition, can imprint signatures in binary neutron star mergers favoring the hypothesis of quark matter in the neutron star cores~\cite{2019JPhG...46k4001A,2019PhRvL.122f1101M,2019PhRvL.122f1102B,2020PhRvL.124q1103W,2021PhRvD.104h3029P}. 

In principle, it would be ideal to describe a neutron star, which consists of a quark phase in its center and a surrounding hadronic phase, in a unified theoretical model for both phases over the whole range of relevant densities. Although such a model is not yet available from a first-principle approach, effective models that incorporate key aspects of QCD provide a viable alternative. From this perspective, for the quark phase, we utilize a typical Nambu–Jona-Lasinio (NJL) model, which effectively captures the dynamics of spontaneous chiral symmetry breaking of QCD~\cite{1992RvMP...64..649K,1994PhR...247..221H,2005PhR...407..205B}, successfully reproduces the spectrum of low-lying mesons~\cite{2005PhR...407..205B,1992RvMP...64..649K,1994PhR...247..221H}, and has been widely used and extended to describe the quark matter in compact star physics, see Refs.~\cite{2003PhRvD..67f5015H,2005PhRvD..72f5020B,2005PhRvD..72f5020B,2019ChPhC..43h4102W,2021PhRvC.104f5201M,2025ApJ...980..231L,2024PhRvD.110a4022X,2025PhRvD.111a4006G,2024arXiv241104064G} as incomplete lists. To improve the description of strongly interacting matter at high baryon chemical potentials, a modified version of the NJL model has been proposed in Ref.~\cite{2019ChPhC..43h4102W}. This extended NJL model incorporates both the original NJL Lagrangian and its Fierz-transformed counterpart, with their respective contributions weighted by the $1-\alpha$ and $\alpha$ parameters. This modified NJL model has been successfully applied in various studies, including the study of the critical endpoint in the QCD phase diagram~\cite{2019PhRvD.100i4012Y,2020ChPhC..44g4104Y}, color superconductivity~\cite{2020PhRvD.102e4028S,2023PhRvD.108d3008Y,2024ApJ...966....3Y}, as well as the properties of quark matter and quark stars~\cite{2020MPLA...3550321W,2022PhRvD.105l3004Y,2024Ap&SS.369...29S}. As an extension to previous studies, we employ this modified NJL-type model to describe the quark phase in the study of hybrid stars and explore the influence of exchange interactions on the implications for the existence of large quark cores in the interior of hybrid stars.  

At lower densities, the bulk properties of strongly interacting matter are significantly influenced by the fact that quarks are confined there. Therefore, we must resort to other models to describe the hadronic matter at low densities currently. Since the NJL model reflects the chiral symmetry of QCD, clearly, it is desirable to have a Lagrangian for the hadronic phase which also respects chiral symmetry, as discussed in Refs.~\cite{1995PhRvC..52.1368F,1997NuPhA.615..441F,1998PhRvC..57.2576P,1999PhRvC..59..411P,1989PhRvD..39.2805D,2001PThPh.106..873J,2008PhRvC..77b5803D}. For this consideration,  the parity doublet model (PDM) is particularly well-suited for this purpose~\cite{1989PhRvD..39.2805D,2001PThPh.106..873J,2008PhRvC..77b5803D,2013PhRvC..87a5804D,2015PhRvC..92b5201M,2017PhRvC..95e9903M,2021PhRvC.103d5205M,2022PhRvC.106f5205G,2023Symm...15..745M,2024PhRvC.109f5807G,2024arXiv240318214G,2024arXiv240805687Y,2024PhRvD.109d1302M,2024PhRvD.110a4018M,2024PhRvC.109d5201E,2024arXiv241016649G,2018PhRvD..98j3021M,2020A&A...643A..82M}, in which the nucleon masses not only have a mass associated with the chiral symmetry breaking, but also a chiral invariant mass $m_0$, which is insensitive to the chiral condensate and the presence of which is manifested by recent lattice QCD simulations~\cite{2015PhRvD..92a4503A,2017JHEP...06..034A,2019PhRvD..99g4503A}. As is well known, a first-order hadron–quark phase transition can be modeled using either the Maxwell or Gibbs construction, depending on the nature of the mixed phase. In particular, the effects of screened Coulomb potentials and surface tension at the phase interface have been extensively studied~\cite{2002PhLB..541...93V,2006nucl.th...5075M,2008PhLB..659..192M,2010JPhG...37b5201B}. However, due to significant uncertainties in model parameters—such as the interface energy—the conclusions remain inconclusive. As noted in Ref.~\cite{2006nucl.th...5075M}, the mixed phase described by the Gibbs construction may, in some cases, be energetically unfavorable and thus excluded from the star. In the present work, we adopt the Maxwell construction for our calculations. In our earlier studies~\cite{2023Symm...15..745M,2024PhRvC.109f5807G,2024arXiv240318214G}, we investigated the crossover scenario assuming zero surface tension. The current study explores the first-order transition under the assumption of infinite surface tension. The Gibbs construction, incorporating finite surface tension, is expected to yield results that lie between these two cases. Employing a Maxwell construction, the deconfinement transition $\mu_{\rm de}$ is associated with the point where both models have the same free energy. Although previous studies have successfully employed the NJL model to describe the quark phase~\cite{1999PhRvC..60b5801S,2007PhLB..654..170K,2012ApJ...759...57L,2016PhRvD..94i4001P,2021PhRvD.103l3020F,2022PhRvC.105d5808C} or color superconducting phase~\cite{2008PhRvD..77f3004P,2012A&A...539A..16B,2024arXiv:2501.00115} in hybrid EOS, predicting hybrid stars with masses exceeding $2\msun$, these works primarily rely on the relativistic mean-field (RMF) model to describe the hadronic phase~\cite{1999PhRvC..60b5801S,2007PhLB..654..170K,2008PhRvD..77f3004P,2012ApJ...759...57L,2012A&A...539A..16B,2016PhRvD..94i4001P,2021PhRvD.103l3020F,2022PhRvC.105d5808C,2024arXiv:2501.00115}, or use meta-models that provide flexibility in modeling the hadronic EOS at low densities~\cite{2020PhRvD.101l3030F}. The Bayesian analyses are also employed to find high chance for the most massive neutron star to host a quark matter core~\cite{2023SciBu..68..913H}, incorporating multi-messenger data of GW170817, PSR J0030 + 0451, PSR J0740 + 6620, and state-of-the-art theoretical progresses, alongside theoretical insights from chiral effective field theory ($\chi$EFT) and perturbative QCD calculations. There is also a hot debate about whether a typical mass of $1.4\msun$ could have a quark core~\cite{1999PhRvC..60b5801S,2020PhRvD.101l3030F,2023PhRvD.108k4028G}. 
We note that, in this study, we perform a comprehensive study of hybrid stars using several parameterizations of a PDM~\cite{1989PhRvD..39.2805D,2001PThPh.106..873J,2008PhRvC..77b5803D,2013PhRvC..87a5804D,2015PhRvC..92b5201M,2017PhRvC..95e9903M,2021PhRvC.103d5205M,2022PhRvC.106f5205G,2023Symm...15..745M,2024PhRvC.109f5807G,2024arXiv240318214G,2024arXiv240805687Y,2024PhRvD.109d1302M,2024PhRvD.110a4018M,2024PhRvC.109d5201E,2024arXiv241016649G} together with a modified NJL model with scalar four-fermion interactions, ’t Hooft six-fermion interactions, and the Fierz transformed interactions~\cite{2019ChPhC..43h4102W,2019PhRvD.100i4012Y,2020ChPhC..44g4104Y,2020PhRvD.102e4028S,2023PhRvD.108d3008Y,2024ApJ...966....3Y,2020MPLA...3550321W,2022PhRvD.105l3004Y,2024Ap&SS.369...29S}. Within this approach, the hadronic and quark degrees of freedom are derived from different theoretical Lagrangians but both respecting chiral symmetry. We aim to systematically explore the possible existence of a large two- or three-flavor quark core in the interior of neutron stars through a first-order deconfinement transition, even for the possible existence of large quark cores in the interior of moderately low mass $\sim 1.4\msun$. This work extends our previous study on the exploration of the first-order phase transition in neutron stars~\cite{2024PhRvC.110d5802G}, in which large quark cores generated by early transition have not been found. We also constrain the maximum mass of the hybrid star within these physical models, and the relation between the phase transition points and the maximum mass of the hybrid stars.

This paper is organized as follows. In Sec.~\ref{Sec: formulalism}, we will briefly present the hadronic EOS employed in this work, and will introduce the modified NJL models for describing the quark matter, including the Fierz transformed interactions. Section~\ref{Sec: Results} discusses the results on hybrid matter EOS and hybrid stars, along with the observational constraints. Our results are summarized in Sec.~\ref{sec:summary}.

\maketitle 
\section{\green{Formulism} }\label{Sec: formulalism}
\subsection{Hadronic matter within PDM model} \label{Sec:hadronic matter}

The nuclear matter EOS within the framework of PDM has been investigated in~\cite{2015PhRvC..92b5201M,2017PhRvC..95e9903M,2021PhRvC.103d5205M,2022PhRvC.106f5205G,2023Symm...15..745M,2024PhRvC.109f5807G,2024arXiv240318214G,2024arXiv240805687Y}. In this approach, ordinary nucleon and its negative-parity excited state are regarded as chiral partners, with their masses becoming degenerate as chiral symmetry is restored at high densities. 
This degenerate mass, known as the chiral-invariant mass $m_0$, is a crucial parameter in this model, which significantly influence the stiffness of the EOS. Specifically, a larger $m_0$ correspond to a weaker $\sigma$ coupling strength, as the nucleon mass is not entirely derived from the $\sigma$ fields. 
This, in turn, leads to weaker $\omega$ field couplings because, at nuclear saturation density, the system requires a balance between the repulsive $\omega$ fields contributions and the attractive $\sigma$ fields interactions. As density increases, the $\sigma$ field strength diminishes while the $\omega$ fields strength grows, disrupting this balance. Consequently, a larger $m_0$ weakens the $\omega$ fields and softens the EOS at supranuclear densities. Typical PDMs incorporate with $\sigma$-$\omega$  mean-field, with some studies also including the isovector scalar meson $a_0(980)$ to model the asymmetric matter configurations in the neutron stars. Nevertheless, as investigated in Ref.~\cite{2023PhRvC.108e5206K}, the inclusion of the $a_0(980)$ has a negligible impact on the properties of neutron stars, resulting in only a slight increase in the radius of less than $1 \rm km$. In this study, we then consider the PDM model with $N_f=2$ and include the vector meson mixing, such as the $\omega^2\rho^2$ interaction, as described in Ref.~\cite{2024PhRvC.109f5807G}.

Following previous studies in Ref.~\cite{2024PhRvC.109f5807G}, the thermodynamic potential of the model is expressed as
\begin{equation}
\begin{aligned}
\Omega_{\mathrm{PDM}}&=V_\sigma-V\left(f_{\pi}\right)\\
&+V_\omega+V_\rho + V_{\omega\rho} +\sum_{i=+,-} \sum_{x=p, n} \Omega_{x} \ ,\label{Eq:Omega PDM}
\end{aligned}
\end{equation}
where $i = +, -$ denote for the parity of the ground state nucleon $N(939)$ and its excited state $N^*(1535)$
The potential $V(\sigma)$, $V_\omega$, $V_\rho$ and $V_{\omega \rho}$ are written as
\begin{equation}
\begin{aligned}
V(\sigma) = -\frac{1}{2}\bar{\mu}^{2}\sigma^{2} &+ \frac{1}{4}\lambda_4 \sigma^4 -\frac{1}{6}\lambda_6\sigma^6 - m_{\pi}^{2} f_{\pi}\sigma\ \ , \\
V_\omega&=-\frac{m_\omega^2}{2} \omega^2 \ ,\\
V_\rho&=-\frac{m_\rho^2}{2} \rho^2\ , \\
V_{\omega \rho}&=-\lambda_{\omega\rho}(g_{\omega NN}\omega)^2(g_{\rho NN}\rho)^2\ ,\label{Eq:PDM potential}
\end{aligned}
\end{equation}
with $\bar{\mu}, \lambda_{4}, \lambda_{6}$ and $\lambda_{\omega\rho}$ are parameters to be determined. Also, the $f_{\pi}$ is the pion decay constant and the kinetic part of the thermodynamic potential $\Omega_x$ is
\begin{equation}
\begin{aligned}
\Omega_x= -2 \int^{k_x^{\pm}} \frac{\mathrm{d}^3 \mathbf{p}}{(2 \pi)^3}\left(\mu_x^*-E_{\mathbf{p}}^i\right),\ \label{Eq: PDM kinetic part} 
\end{aligned}
\end{equation}
where $E_{{\bf p}}^{i} = \sqrt{{\bf p}^{2} + m_{\pm}^{2}}$ represents the energy of the corresponding nucleon with mass $m_{\pm}$ and momentum ${\bf p}$, and $k_{x}^{\pm}=\sqrt{(\mu^{*}_{x})^2-m_{\pm}^2}$ defines the Fermi momentum for the relevant particle, with $\mu_x^{*}$ being the effective chemical potential. In our calculations, we employ the no-sea approximation, which assumes that the Dirac sea structure remains identical for both vacuum and medium conditions.

The masses of positive- and negative-parity chiral partners are expressed as
\begin{equation}
\begin{aligned}
m_{ \pm}=\frac{1}{2}\left[\sqrt{\left(g_1+g_2\right)^2 \sigma^2+4\left(m_0\right)^2} \mp\left(g_1 - g_2\right) \sigma\right]\ ,\label{Eq: PDM mass}
\end{aligned}
\end{equation}
 For a given chiral invariant mass $\mathrm{m}_0$, the parameters $\mathrm{g}_1$ and $\mathrm{g}_2$ are determined by the corresponding vacuum masses, $m_N=939 \mathrm{MeV},  m_{N^*}=1500 \mathrm{MeV}$. The effective chemical potentials for nucleons and their chiral partners are given by
\begin{equation}
\begin{aligned}
\mu_p=\mu_p^{*}&=\mu_Q+\mu_B-g_{\omega NN} \omega- \frac{1}{2}g_{\rho NN} \rho \ , \\
\mu_n=\mu_n^{*}&=\mu_B - g_{\omega NN} \omega+ \frac{1}{2}g_{\rho NN} \rho \ .\label{Eq: PDM chemicalPoten}
\end{aligned}
\end{equation}
 
The complete thermodynamic potential for hadronic matter in neutron stars includes lepton contributions
\begin{equation}
\begin{aligned}
\Omega_{\mathrm{H}}=\Omega_{\mathrm{PDM}} + \Omega_{e}\ ,
\end{aligned}
\end{equation}
where $\Omega_{e}$ represents the thermodynamic potential for electrons
\begin{equation}
\begin{aligned}
\Omega_{e}=-2 \int^{k_F} \frac{\mathrm{d}^3 \mathbf{p}}{(2 \pi)^3}\left(\mu_l-E_{\mathbf{p}}^l\right)\ ,
\end{aligned}
\end{equation}
Finally, we have the pressure in hadronic matter as
\begin{equation}
P_{\mathrm{H}}=-\Omega_{\mathrm{H}}\ .
\end{equation}
Using parameter sets established in Ref.~\cite{2024PhRvC.109f5807G}, which were calibrated to reproduce the normal nuclear matter properties, we calculate the hadronic phase EOS for various values of the chiral invariant mass $m_0$.


\subsection{Quark matter within modified NJL model}\label{Sec: NJL model}
In this section, we introduce the modified NJL model with exchange interacting channels to describe the effective interactions between quarks. 
\subsubsection{two-flavor modified NJL model}\label{Sec: 2f NJL model}
The Lagrangian of the two-flavor NJL model reads:
\begin{equation}
\mathcal{L}_{\mathrm{NJL}}^{~2f} =\mathcal{L}_{0}+G\left[(\bar{\psi} \psi)^{2}+\left(\bar{\psi} i \gamma^{5} \tau \psi\right)^{2}\right] \ , \label{eq1}
\end{equation}
in which $\mathcal{L}_{0} =\bar{\psi}\left(i \gamma^{\mu}\partial_{\mu} -m +\mu \gamma^{0} \right)\psi$ is the relativistic free field which describes the propagation of non-interacting fermions. $\psi$ is the quark field operator with color, flavor, and Dirac indices. $G$ is the four-fermion interaction coupling constants. $\mu$ is the flavor-dependent quark chemical potential. $m$ is the diagonal mass matrix for quarks in flavor space $m=\mathrm{diag}(m_u,m_d)$, which contains the small current quark masses and introduces a small explicit chiral symmetry breaking. Here, we take $m_u=m_d$.

For further considering the effect of a rearrangement of fermion field operators, we apply the Fierz transformation to the interaction terms in the NJL models, as discussed in Ref.~\cite{2019ChPhC..43h4102W,2019PhRvD.100i4012Y,2020ChPhC..44g4104Y,2020PhRvD.102e4028S,2023PhRvD.108d3008Y,2024ApJ...966....3Y,2020MPLA...3550321W,2022PhRvD.105l3004Y,2024Ap&SS.369...29S}. As a purely technical device to examine the exchange channels influence that occur in quartic products at the same space-time point~\cite{1992RvMP...64..649K,2005PhR...407..205B}, the Fierz identity of the four-fermion interactions in the two-flavor NJL model is
\begin{equation}
\begin{aligned}
\mathcal{F}(\mathcal{L}_{\sigma}^{4})=& \frac{G}{8 N_{c}}\left[2(\bar{\psi} \psi)^{2}+2\left(\bar{\psi} i \gamma^{5} \tau \psi\right)^{2}-2(\bar{\psi} \tau \psi)^{2}\right.\\
&-2\left(\bar{\psi} i \gamma^{5} \psi\right)^{2}-4\left(\bar{\psi} \gamma^{\mu} \psi\right)^{2}-4\left(\bar{\psi} i \gamma^{\mu} \gamma^{5} \psi\right)^{2} \\
&\left.+\left(\bar{\psi} \sigma^{\mu \nu} \psi\right)^{2}-\left(\bar{\psi} \sigma^{\mu \nu} \tau \psi\right)^{2}\right] \ , \label{eq5}
\end{aligned}
\end{equation}
Here, $N_c$ is the number of color which is given by $N_c=3$ and we only consider the contribution of color singlet terms for simplicity. One can see that, in Eq.~(\ref{eq5}), the Fierz transformed Lagrangian contains not only the scalar and pseudoscalar interactions, but also vector and axialvector interaction channels.

Due to the mathematical equality between the original interactions and Fierz transformed interactions, we can combine them using a weighting factor $\alpha$. The factor $\alpha$ reflects the competition between the original interaction channels and the exchange interaction channels.
Then the effective Lagrangian becomes
\begin{equation}
\mathcal{L}_{\rm eff}^{~2f}=\bar{\psi}(i \gamma^{\mu}\partial_{\mu} -m+\mu\gamma^{0}) \psi +(1-\alpha)\mathcal{L}_{\mathrm{int}}^{~2f}+\alpha \mathcal{F}(\mathcal{L}_{\mathrm{int}}^{~2f}) \ .\label{eq7}
\end{equation}

Under the mean-field approximation, the mass gap equation and the effective chemical potential can be obtained as follows:
\begin{equation}
\begin{aligned}
M=&m -2\left[(1-\alpha) +\frac{\alpha }{12} \right]G \sum_{f=u, d}\sigma_f \\
 =&m-2 G' \sum_{f=u, d}\sigma_f \ ,\\
 \mu^{*} =&\mu-\frac{\alpha}{3} G\sum_{f=u, d}\rho_f\ ,
\end{aligned} \label{eq: 2f gap Eq}
\end{equation}
where $G'=(12-11\alpha)G/12$. The quark condensate $\langle\bar{\psi} \psi\rangle$ and quark number density $\left\langle\psi^{+} \psi\right\rangle$ are denoted as $\sigma$ and $\rho$, respectively, which are the average values of operaters, $\bar{\psi} \psi$ and $\psi^{+} \psi$, in the ground state. 

\subsubsection{(2+1)-flavor modified NJL model}\label{Sec: 3f NJL model}
The Lagrangian density of the (2+1)-flavor NJL model is given by  
\begin{equation}
\begin{aligned} 
\mathcal{L}_{\mathrm{NJL}}^{~3f} &=\mathcal{L}_{0}+\mathcal{L}_{\mathrm{int}}^{~3f} \ , \\
\mathcal{L}_{\mathrm{int}}^{~3f} &=  \mathcal{L}_{\sigma}^{4}+ \mathcal{L}_{\sigma}^{6}  \ ,
\label{eqNJL}
\end{aligned}
\end{equation}  
where $\mathcal{L}_{\sigma}^{4}$ and $\mathcal{L}_{\sigma}^{6}$ represent the four-fermion and six-fermion interaction terms, respectively. These interaction terms are expressed as  
\begin{equation}
\begin{aligned}
\mathcal{L}_{\sigma}^{4} &=\sum_{i=0}^{8} G\left[\left(\bar{\psi} \lambda_{i} \psi\right)^{2}+\left(\bar{\psi} i\gamma^{5}\lambda_{i} \psi\right)^{2}\right] \ ,\\ 
\mathcal{L}_{\sigma}^{6} &= -K\left(\operatorname{det}\left[\bar{\psi}\left(1+\gamma^{5}\right) \psi\right]+\operatorname{det}\left[\bar{\psi}\left(1-\gamma^{5}\right) \psi\right]\right) \ . \label{eqNJLint}
\end{aligned}
\end{equation}  
Here, $G$ and $K$ denote the coupling constants for the four-fermion and six-fermion interactions, respectively. The matrices $\lambda_{i}\; (i=1 \rightarrow 8)$ correspond to the Gell-Mann matrices in flavor space, while $\lambda_{0} = \sqrt{2/3}\; I_{0}$, with $I_{0}$ being the identity matrix.  
Applying the Fierz transformation to the four-fermion scalar and pseudoscalar interaction terms, considering only the contributions from color-singlet channels, yields  
\begin{equation}
\mathcal{F}(\mathcal{L}_{\sigma}^{4})=-\frac{G}{2} \left[\left(\bar{\psi} \gamma_{\mu} \lambda_{i}^{0} \psi\right)^{2}-\left(\bar{\psi} \gamma_{\mu} \gamma_{5} \lambda_{i}^{0} \psi\right)^{2}\right] \ .\label{eqFierzscalar}
\end{equation}  
Since the Fierz transformation of the six-fermion interaction is defined to preserve invariance under all possible permutations of the quark spinors $\psi$ appearing in the interaction~\cite{1992RvMP...64..649K}, the six-fermion term remains unchanged:  
\begin{equation}
\mathcal{F}(\mathcal{L}_{\sigma}^{6})=\mathcal{L}_{\sigma}^{6} \ .\label{eqFierzSix}
\end{equation}  
Thus, the effective Lagrangian takes the form  
\begin{equation}
\mathcal{L}_{\rm eff}^{~3f}=\bar{\psi}(i \gamma^{\mu}\partial_{\mu} -m+\mu\gamma^{0}) \psi +(1-\alpha)\mathcal{L}_{\mathrm{int}}^{~3f}+\alpha \mathcal{F}(\mathcal{L}_{\mathrm{int}}^{~3f}) \ .\label{eqNJLeff}
\end{equation}  

In the mean-field approximation, the mass gap equations and the effective chemical potential $\mu_{f}^{*}$ for a given flavor $f$ are given by  
\begin{equation}
\begin{aligned}
M_{f}&= m_{f}-4 (1-\alpha) G \sigma_{f}+2 K \sigma_{j} \sigma_{k}\\
&= m_{f}-4G'\sigma_{f}+2 K \sigma_{j} \sigma_{k} , \\
\mu_{f}^{*}&=\mu_{f} -\frac{2}{3}\alpha G\sum_{f^{\prime}=u, d, s}\rho_{f^{\prime}}\ ,\label{eq:Effmass}
\end{aligned}
\end{equation}  
where we define $G' = (1-\alpha)G$, and $f, j, k $ represent even permutations of \( u, d, s \). From Eq.~(\ref{eq:Effmass}), it is evident that the introduction of the Fierz-transformed interactions contributes to the effective chemical potential and modifies the gap equation. 


\begin{figure*}
\centering
{\includegraphics[width=0.49\textwidth]{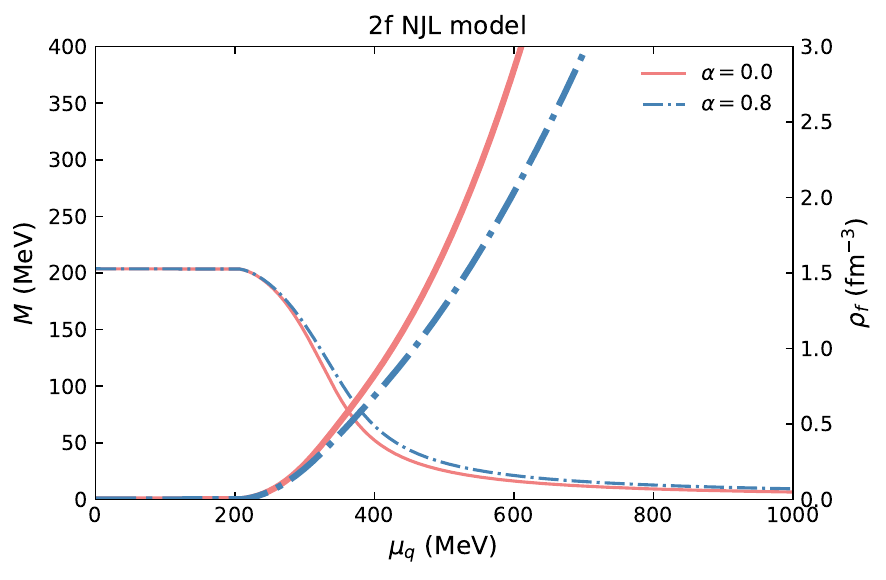}}
{\includegraphics[width=0.49\textwidth]{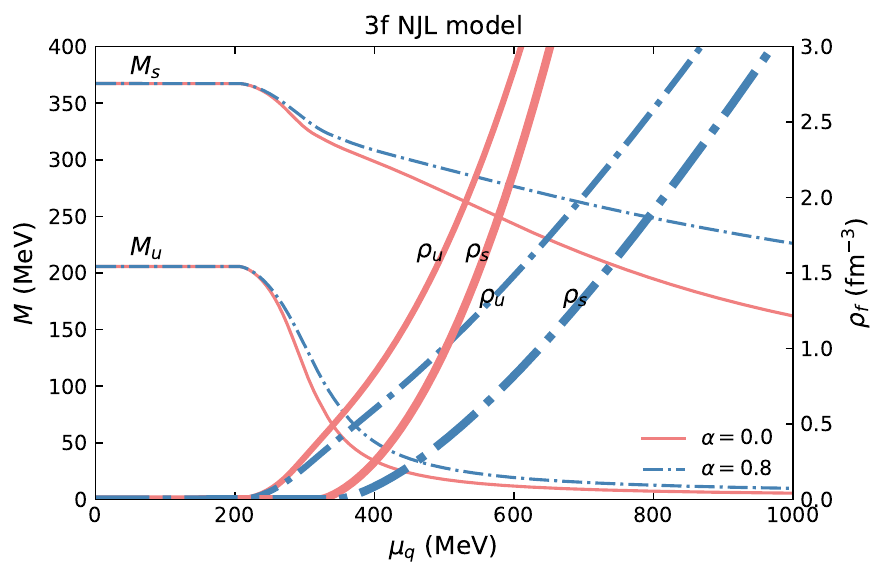}}
\caption{The dynamical quark mass $M$ of $u$, $d$ and $s$ quarks versus the quark chemical potential $\mu_q$, as well as quark number density $\rho_f$ ($f=u, d, s$) versus $\mu_q$ for two-flavor and (2+1)-flavor NJL modified NJL models with $\alpha=0.0$ and $\alpha=0.8$. The red curves represent the results for $\alpha=0.0$, while the dot-dashed blue curves exhibit the results for $\alpha=0.8$.
}
\label{fig:mass_rho}
\end{figure*}
\subsubsection{quark condensate and quark number density at zero temperature}
In the following, we focus on deriving the key quantities—quark condensate and quark number density—while presenting the regularization procedure we employed. In Euclidean space, introducing a finite chemical potential at zero temperature is equivalent to performing a transformation~\cite{1994PhR...247..221H,2005PhR...407..205B,2005PhRvC..71a5205Z}: $p_{4} \rightarrow p_{4}+i \mu_{f}^{*}$. By first integrating over $p_4$ and applying proper-time regularization,
\begin{equation}
\begin{aligned}
\frac{1}{A^{n}} &= \frac{1}{(n-1) !} \int_{0}^{\infty} \mathrm{d} \tau \, \tau^{n-1} e^{-\tau A} \\
&\stackrel{\text { UV cutoff }}{\longrightarrow} \frac{1}{(n-1) !} \int_{\tau_{\mathrm{UV}}}^{\infty} \mathrm{d} \tau \, \tau^{n-1} e^{-\tau A}\ ,
\label{eq:NJL PRT}
\end{aligned}
\end{equation}
where the lower cutoff $\tau_{\rm UV} = 1/\Lambda_{\rm UV}^2$ suppresses high-frequency contributions~\cite{1994PhR...247..221H,2005PhR...407..205B} with parameter $\lambda_{\rm UV}$ related to the ultraviolet, one obtains the quark condensate as
\begin{equation}
\begin{aligned}
&\sigma_{f} = -N_{\mathrm{c}} \int_{-\infty}^{+\infty} \frac{\mathrm{d}^{4} p^{\mathrm{E}}}{(2 \pi)^{4}} \frac{4  M_{f}}{\left(  p_{4}+ i\mu_{f}^{*} \right)^{2}+ p^{2}+M_{f}^{2}} \\
&= 
\begin{cases} 
-\frac{3 M_{f}}{\pi^{2}} \int_{\sqrt{\mu_{f}^{* 2}-M_{f}^{2}}}^{+\infty} \mathrm{d} p \frac{\left[1-\operatorname{Erf}\left(\sqrt{M_{f}^{2}+p^{2}} \sqrt{\tau_{\mathrm{UV}}}\right)\right] p^{2}}{\sqrt{M_{f}^{2}+p^{2}}}\ ,  & M_{f}<\mu_{f}^{*} \\ 
\frac{3 M_{f}}{4 \pi^{2}}\left[M_{f}^{2} \Gamma\left(0, M_{f}^{2}\tau_{\mathrm{UV}} \right) -\frac{e^{-M_{f}^{2} \tau_{\mathrm{Uv}}}}{\tau_{\mathrm{UV}}}\right]\ , & M_{f}>\mu_{f}^{*}
\end{cases} 
\label{eqNJLsigma}
\end{aligned}
\end{equation}
here, $\Gamma(a, z) = \int_{z}^{+\infty} \mathrm{d} t \; t^{a-1} e^{-t}$ is the gamma function, and $\operatorname{Erf}(x)$ is the error function. 

At zero temperature and finite chemical potential, the quark number density is given by  
\begin{equation}
\begin{aligned}
\rho_{f} &=\left\langle\psi^{+} \psi\right\rangle_{f} =-\int \frac{\mathrm{d}^{4} p}{(2 \pi)^{4}} \operatorname{Tr}\left[i S_{f}\left(p^{2}\right) \gamma_{0}\right] \\
&=2 N_{\mathrm{c}} \int \frac{\mathrm{d}^{3} p}{(2 \pi)^{3}} \theta\left(\mu_{f}^{*}-\sqrt{p^{2}+M_{f}^{2}}\right) \\
&= 
\begin{cases} 
\frac{1}{\pi^{2}}\left(\sqrt{\mu_{f}^{* 2}-M_{f}^{2}}\right)^{3}, & \mu_{f}^{*}>M_{f} \\ 
0\ . & \mu_{f}^{*}<M_{f} 
\end{cases}
\label{eqNJLQN}
\end{aligned}
\end{equation}
Equation~(\ref{eqNJLQN}) explicitly indicates that the quark number density for flavor $f$ becomes nonzero only when the effective quark chemical potential $\mu^{*}_f$ exceeds a certain threshold.

Before performing calculations, we should first determine the model parameters. At zero temperature and quark chemical potential, apart from $\alpha$, the determination of model parameters follows the standard NJL model~\cite{1994PhR...247..221H}. After setting the up and down quark masses to equal values, the remaining parameters $m_s$, $\Lambda_{\mathrm{UV}}$, $G'$, $K$ are chosen to reproduce experimental values of the pion decay constant and meson masses: $f_{\pi}=92$~MeV, $M_{\pi}=135$ MeV, $M_{K^{0}}=495$~MeV, $M_{\eta}=548$~MeV, and $M_{\eta^{\prime}}=958$~MeV. For the (2+1)-flavor NJL model, the parameters are chosen as $m_u=3.4\mev$, $m_s=104\mev$, $\Lambda=1330\mev$, $G'=1.51\times10^{-6}\mev^{-2}$, $K=2.75\times10^{-14}\mev^{-5}$. For the two-flavor NJL model, we adopt $m_u=3.3\mev$, $\Lambda=1330\mev$, $G'=2.028\times10^{-6}\mev^{-2}$~\cite{2022PhRvD.105l3004Y}.

By solving the mass gap equations in Eq.~(\ref{eq: 2f gap Eq}) and Eq.~(\ref{eq:Effmass}) for the modified NJL models, we can obtain the dynamical quark masses as functions of the quark chemical potentials, as shown in Fig.~\ref{fig:mass_rho}. When $\mu_f^* < M_f$, the dynamical quark masses remain at their vacuum values, indicating strong interactions and quark confinement. With the increasing of the quark chemical potential, the dynamical quark masses decrease for $\mu_f^* > M_f$, and simultaneously, the quark number densities become nonzero, as illustrated in the right panel of Fig.~\ref{fig:mass_rho}. Increasing $\alpha$ strengthens the vector interactions from the Fierz-transformed channels in Eq.~(\ref{eqFierzscalar}), which causes the dynamical mass to decrease more slowly and results in a stiffer EOS compared to the original NJL model. As will be discussed in detail in Section~\ref{Sec: Results}, these changes will influence the deconfinement transition and the hybrid EOS.

\subsection{QCD vacuum pressure as free parameter}  \label{Subsec: QCD vacuum pressure}
At finite chemical potential and zero temperature, the pressure for quark matter can be strictly proved with the functional path integrals of QCD ~\cite{2008PhRvD..78e4001Z,2008IJMPA..23.3591Z}:
\begin{equation}
P(\mu; M)=P(\mu=0; M)+\int_{0}^{\mu} d \mu^{\prime} \rho\left(\mu^{\prime}\right) , \label{eq: njLPressure}
\end{equation}
where $M$ represents a solution to the previously discussed gap equation. The energy density and pressure of the system are related by the thermodynamic equation:
\begin{equation}
\varepsilon = -P + \sum_{i=u, d, s, e} \mu_i \rho_i(\mu_i) \ , \label{eq:NJL eos}
\end{equation}
with the conditions of $\beta$-stability and charge neutrality. Given the non-perturbative difficulty of directly calculating the vacuum pressure $P(\mu=0; M)$ from first-principles QCD, effective QCD models are often employed. In NJL-type models, people usually choose the trivial vacuum as $P(\mu=0; m)$, and evaluate the vacuum pressure difference between the trivial vacuum $P(\mu=0; m)$ and the spontaneous symmetry breaking non-trivial vacuum $P(\mu=0;M)$ to determine the vacuum pressure, as extensively discussed in Ref.~\cite{2022PhRvD.105l3004Y}. Nevertheless, because of the lack of confinement at vanishing density, this procedure to determine the vacuum pressure is unsatisfactory. Therefore, in this study, we take $P(\mu=0; M)$ as a phenomenological free parameter corresponding to $-B$~\cite{2012ApJ...759...57L,2022PhRvD.105l3004Y,2024ApJ...966....3Y,2024arXiv240805687Y}, thereby preserving quark confinement.

\begin{figure}
\centering
\includegraphics[width=0.49\textwidth]{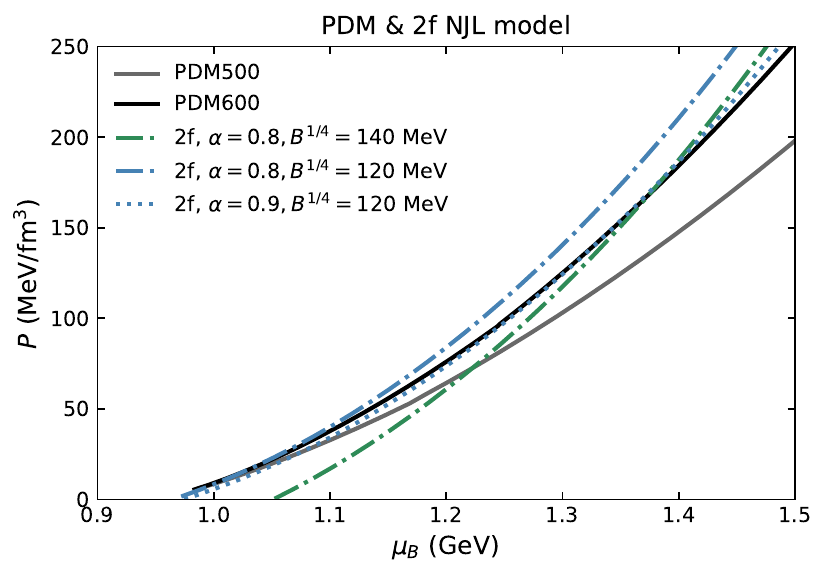}
{\includegraphics[width=0.49\textwidth]{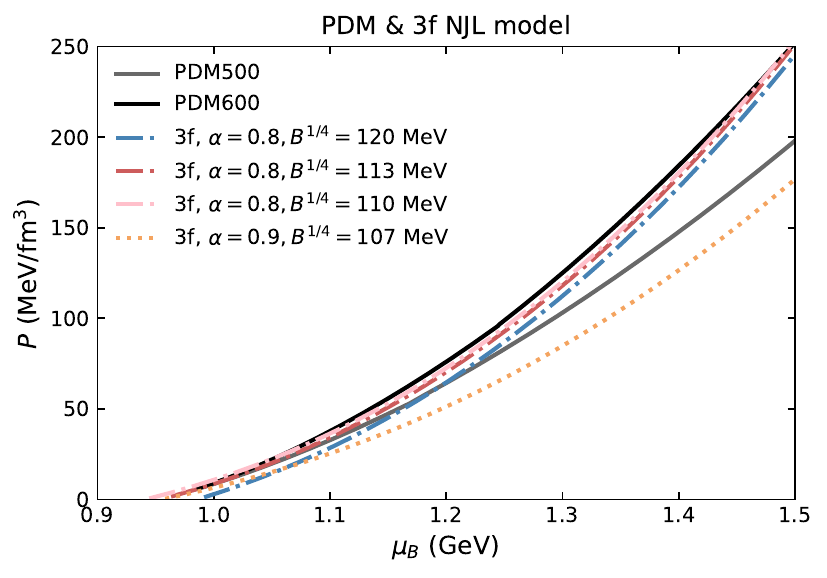}}
\caption{Pressure $P$ as a function of baryon chemical potential $\mu_B$ for hadronic matter and quark matter. In each panel, the PDM calculations for hadronic matter with chiral invariant masses of $m_0 = 500~\mathrm{MeV}$ and $m_0 = 600~\mathrm{MeV}$ are shown as gray and black curves, respectively. For quark matter, curves with the same style represent results for the same value of $\alpha$, while different colors correspond to different vacuum pressures. For instance, in the left panel, the blue curves denote results from the two-flavor modified NJL model with $B^{1/4} = 120~\mathrm{MeV}$ at various values of $\alpha=0.8, \ 0.9$.
}\label{fig:Pmu}
\end{figure}

\section{EOS and Hybrid star structure with large quark core}\label{Sec: Results} 
In this section, we solve the Tolman-Oppenheimer-Volkoff (TOV) equations~\cite{1939PhRv...55..364T,1939PhRv...55..374O} for spherically symmetric, static stars to explore the impact of the exchange channels weighted by the parameter $\alpha$, as well as the crucial role of the vacuum pressure $B$ in determining the existence of a large quark core and the maximum mass of hybrid stars. For the neutron star crust, we use the Baym-Pethick-Sutherland (BPS) EOS~\cite{1971ApJ...170..299B}, where the energy density ranges from $1.0317 \times 10^4$~$\rm g/cm^3$ to $4.3 \times 10^{11}$~$\rm g/cm^3$. The outer core is modeled by the PDM, while the potential quark core is described by the modified NJL model.

\subsection{Pressure versus baryon chemical potential}\label{Sec: Hybrid star EOS}
We illustrate the pressure $P$ as functions of baryon chemical potential $\mu_B$ for the hadronic phase and quark phase in Fig.~\ref{fig:Pmu}. In this figure, $\mathrm{PDM500}$ indicates the PDM with chiral invariant mass $m_0$ for $500\mev$. In the PDM framework, a larger chiral invariant mass $m_0$ results in a softer EOS in the hadronic region. Our previous studies~\cite{2015PhRvC..92b5201M,2017PhRvC..95e9903M,2021PhRvC.103d5205M,2021PhRvC.104f5201M,2022PhRvC.106f5205G,2023Symm...15..745M,2024PhRvC.109f5807G} have shown that pure hadronic EOS with $m_0=700\mev$ cannot support the observed massive neutron stars with $2M_\odot$. 
Additionally, the presence of a first-order phase transition further softens the EOS, making it even more challenging to comply with $2 M_{\odot}$ constraint~\cite{1999PhRvC..60b5801S,2007PhLB..654..170K,2008PhRvD..77f3004P,2012ApJ...759...57L,2012A&A...539A..16B,2016PhRvD..94i4001P,2021PhRvD.103l3020F,2022PhRvC.105d5808C,2024arXiv:2501.00115}. Therefore, we focus on the PDM with $m_0=500\mev$ and $m_0=600\mev$ contributed to relatively stiff EOS for our present study.

In both panels of Fig.~\ref{fig:Pmu}, the dot-dashed curves depict the results from NJL models for $\alpha=0.8$, with different colors denoting different vacuum pressures $B^{1/4}$. The blues curves illustrate the results from NJL models for the same vacuum pressure at $B^{1/4}=120 \mev$, with various curve styles representing different values of $\alpha$. Within this modified NJL framework, larger values of $\alpha$ correspond to a stiffer quark matter EOS, attributed to the enhanced repulsive interactions in the exchange channels which can be found by combing Eq.~(\ref{eq5}) and Eq.~(\ref{eq7}) [Eq.~(\ref{eqNJLint}) and Eq.~(\ref{eqFierzscalar})] for two-flavor model and (2+1)-flavor model, respectively. Furthermore, a higher vacuum pressure $-B$ in the NJL model significantly stiffens the quark matter EOS, which can be clearly understood from its definition: a higher $-B$ increases the pressure for a given energy density, making the EOS stiffer.
Thermodynamic stability requires that, for a given baryon chemical potential $\mu_B$, the phase with the higher pressure $P$ is more stable~\cite{1985tait.book.....C}. As a result, the pressure-versus-baryon-chemical-potential $P(\mu_B)$ relations of the two phases must intersect at least once, with the point of intersection defining the deconfinement chemical potential, $\mu_{\rm de}$. Under the Maxwell construction, both the chemical potential $\mu_B$ and the pressure $P$ remain continuous at the phase transition point. In general, the $P(\mu_B)$ curves of the two phases can intersect multiple times. In the scenario with a single intersection, where stable hadronic matter exists at low densities, one inevitably finds that $\mu_{\rm hadron} < \mu_{\rm quark}$ at $P=0$. This leads to the energy per baryon satisfying $(\epsilon/n_B)_{\rm hadron} < (\epsilon/n_B)_{\rm quark}$ at $P=0$, indicating an unstable quark matter state, as derived from the gross thermodynamic properties of the system, as discussed in Ref.~\cite{2024arXiv241019678B}. This scenario supports the neutron and hybrid star models. However, if the quark matter phase is more stable at low densities, with $\mu_{\rm hadron} > \mu_{\rm quark}$ at $P=0$, an intersection will occur at $\mu_{\rm low}$. The QCD theory tells us that the quark matter phase will anticipate in the system at very high $\mu_B$. Consequently, another intersection should appear at relatively high densities $\mu_{\rm high}$ compared to $\mu_{\rm low}$. In this case, quark matter has a lower energy per baryon than hadronic matter at $P=0$, i.e., $(\epsilon/n_B)_{\rm hadron} > (\epsilon/n_B)_{\rm quark}$, which supports the quark star hypothesis~\cite{1971PhRvD...4.1601B,1984PhRvD..30..272W,2018PhRvL.120v2001H}. In the following, we will show that further incorporating the details of the microscopic physics within our models will also result in different behaviors in the $P(\mu_B)$ intersections.

For two-flavor quark matter, the vacuum pressure must exceed $B^{1/4}=115~\mathrm{MeV}$ to ensure that there exists an intersection with the hadronic matter EOS. At a fixed vacuum pressure of $B^{1/4}=120~\mathrm{MeV}$ in the modified NJL model, coupled with the PDM model for $m_0=500~\mathrm{MeV}$, strong repulsive interactions in the exchange channels at large $\alpha$ values (e.g., $\alpha=0.8$ and $\alpha=0.9$) induce sharp phase transitions at low baryon chemical potentials, occurring at $\mu_B = 1.0~\mathrm{GeV}$ and $\mu_B = 1.08~\mathrm{GeV}$, respectively, as shown in Table.~\ref{table:PDM_2f_NJL}. These sharp transitions may result in the formation of a substantial quark core within hybrid stars, which will be shown in Sec.~\ref{Sec: Hybrid star structure}. Increasing the exchange interacting channels, which enhances the repulsive interaction, shifts the $\mu_{\rm de}$ to higher baryon chemical potentials. 
In constructing the PDM600 model with various two-flavor quark matter EOSs, the parameter space in achieving a significant quark core larger than $1\msun$ in the interior of the hybrid star with $M_{\rm TOV}$ exceeding $2\msun$ is narrowed. As shown in
the left panel of Fig.~\ref{fig:Pmu}, for two-flavor quark matter with $\alpha = 0.9$ at $B^{1/4} = 120~\mathrm{MeV}$, the EOS properties of the PDM600 model and the two-flavor quark matter become very similar. This similarity results in a large $\mu_{\rm de}$ when the vacuum pressure is slightly increased, further reducing the possibility of a low baryon chemical potential phase transition. Nonetheless, the relatively stiff quark matter EOS for $\alpha=0.7, \ 0.8$, can realize a large quark core larger than $1\msun$ with $M_{\rm TOV}$ around $2\msun$.

The $P(\mu_B)$ curves for (2+1)-flavor quark matter exhibit more complex behavior compared to the two-flavor case. As illustrated in the right panel of Fig.~\ref{fig:Pmu}, results from the (2+1)-flavor modified NJL model are presented for different parameter sets of $\alpha$ and $B^{1/4}$. For $B^{1/4} = 120~\mathrm{MeV}$ at $\alpha=0.8$, there is an intersection with nuclear EOS for PDM600 at a high baryon chemical potential around $\mu_B = 1.565~\mathrm{GeV}$. Reducing the vacuum pressure shifts this intersection to lower baryon chemical potential. When the vacuum pressure drops below $113~\mathrm{MeV}$, such as $B^{1/4} = 110~\mathrm{MeV}$, the $P(\mu_B)$ curves for hadronic and quark matter intersect twice: once at $\mu_B \sim 1.05~\mathrm{GeV}$ and again at $\mu_B \sim 1.5~\mathrm{GeV}$. This suggests that the vacuum pressure in the NJL model influences both the energy per baryon at $P=0$ and the characteristics of the $P(\mu_B)$ curve intersections. 
Another noteworthy feature arises when $\alpha = 0.9$, where the increased stiffness of the (2+1)-flavor quark matter EOS causes the $P(\mu_B)$ curves of the two phases to intersect near $P = 0$. However, when the vacuum pressure exceeds $B^{1/4} = 107~\mathrm{MeV}$, no intersection occurs, indicating that quark matter cannot achieve thermodynamic stability at lower baryon chemical potentials. Conversely, for vacuum pressures below $B^{1/4} = 107~\mathrm{MeV}$, quark matter becomes the more stable phase at lower baryon chemical potentials. In this case, we expect an additional intersection of the $P(\mu_B)$ curves at high chemical potential, ensuring that the quark phase remains stable at very high densities. However, no such second intersection appears. Therefore, we exclude the quark phase within the NJL model for $\alpha=0.9$, as it does not meet the required conditions. As we can see, both the non-perturbative influence of the exchange channel and the vacuum bag constant play crucial roles in determining the stable phase across different density ranges.
For simplicity, we primarily focus on scenarios involving a single intersection, corresponding to stable hadronic matter at low chemical potentials and a traditional phase transition to quark matter at higher densities. We note that, for (2+1)-flavor quark matter EOS, achieving a significant quark core larger than $1\msun$ in the interior of the hybrid star with $M_{\rm TOV}$ exceeding $2\msun$ becomes highly challenging. Because the quark matter EOS must be sufficiently stiff, characterized by a large $\alpha$, to support a massive star with $2~M_\odot$, the vacuum pressure must simultaneously remain a relatively not too large value to allow a phase transition at low baryon chemical potentials. Satisfying both conditions concurrently is particularly difficult for the strange quark matter EOS when constructing the PDM. This will be discussed in detail in Sec.~\ref{Sec: Hybrid star structure}.

\begin{figure}
\centering
{\includegraphics[width=0.48\textwidth]{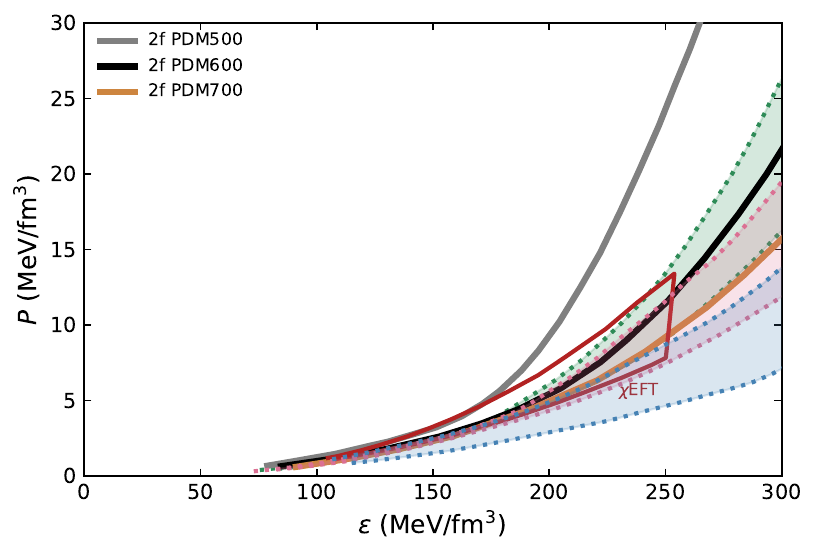}}
\caption{The pressure \( P \) as functions of energy density \( \epsilon \), derived from PDM and predicted from $\chi$EFT, are shown. The gray, black, and sandy brown lines represent hadronic EOSs derived from the PDM with \( m_0 = 500~\mathrm{MeV} \), \( m_0 = 600~\mathrm{MeV} \), and  \( m_0 = 700~\mathrm{MeV} \), respectively. The red band denotes the \( 1\sigma \) uncertainty in \(\chi\)EFT~\cite{2021PhRvC.103d5808D,2023PhRvL.130g2701K}. The other data-driven results incorporating constraints from \(\chi\)EFT within the \( 1\sigma \) uncertainty are also shown together: the blue and violet-red bands represent the results of Bayesian analyses~\cite{2021ApJ...918L..29R} and EOSs inferred from data-driven deep learning methods~\cite{2021JHEP...03..273F}, respectively. The green band shows the Bayesian analyses results mainly constrained by neutron star observations~\cite{Ozel:2015fia,Bogdanov:2016nle}.}
\label{fig:eosEFT} 
\end{figure}

As is well established, $\chi$EFT serves as a powerful framework for performing microscopic calculations of nuclear matter properties at densities up to approximately $2~\rho_0$~\cite{2019PhRvL.122d2501D,2021ARNPS..71..403D,2023PhRvL.130g2701K,2023PhRvL.130i1404F}. Within its domain of applicability, the theory provides a systematic expansion for two-nucleon and multi-nucleon interactions that remain consistent with the symmetries of low-energy QCD. A major advantage over phenomenological approaches is that $\chi$EFT allows for the quantification of theoretical uncertainties through order-by-order convergence analysis of its expansion. To date, $\rm N^3LO$ $\chi$EFT calculations have been employed in studies of nuclear properties~\cite{2019PhRvL.122d2501D,2023PhRvL.130g2701K} as well as in the analysis of neutron star matter~\cite{2021PhRvC.103d5808D}. In Fig.~\ref{fig:eosEFT}, we compare the EOS derived from the PDM (with $m_0 = 500,~600,~700\mev$) against the theoretical window predicted by $\rm N^3LO$ $\chi$EFT with $1\sigma$ uncertainty shown as the red band~\cite{2019PhRvL.122d2501D,2021PhRvC.103d5808D,2023PhRvL.130g2701K}, in which the uncertainty stems from truncation errors and the limitations of many-body perturbation theory approximation. We also present data-driven results that incorporate $\chi$EFT constraints, shown as blue and violet-red bands, representing the Bayesian analysis from Ref.~\cite{2021ApJ...918L..29R} and the data-driven deep learning approach from Ref.~\cite{2021JHEP...03..273F}, respectively. The green band corresponds to results primarily constrained by neutron star observations~\cite{Ozel:2015fia,Bogdanov:2016nle}. Figure~\ref{fig:eosEFT} reveals that a larger invariant mass, such as $m_0 = 600\mev$, yields relatively better agreement with $\chi$EFT predictions compared to $m_0 = 500\mev$, as the author found in Ref.~\cite{2025arXiv250521970G}. 
To further constrain the EOSs derived from the PDM and NJL models, it is essential to incorporate all available and well-controlled theoretical inputs relevant to the densities encountered in compact stars. For example, Ref.~\cite{2022PhRvL.128t2701K} constructed a constrained EOS in the \(\epsilon\)-\(P\) plane by interpolating between state-of-the-art low-density calculations from \(\chi\)EFT near \(0.176\,\mathrm{fm}^{-3}\) and perturbative QCD results at high densities around \(\sim 40\,\rho_0\). This interpolation respects both causality and thermodynamic consistency. Although it excludes previously unconstrained regions, the remaining allowed domain remains sufficiently broad to accommodate our effective field theory results at intermediate densities. The pressure and energy densities predicted by the hadronic and hybrid EOSs based on the PDM and NJL models do not exceed \(500\,\mathrm{MeV/fm^3}\) in the cores of compact stars. These values lie comfortably within the permitted region (green contour in Fig.~4 in Ref.~\cite{2022PhRvL.128t2701K}), where the pressure and energy density can reach up to \(10^4\,\mathrm{MeV/fm^3}\).

\begin{figure*}[htbp]\centering
\includegraphics[width=0.9\hsize]{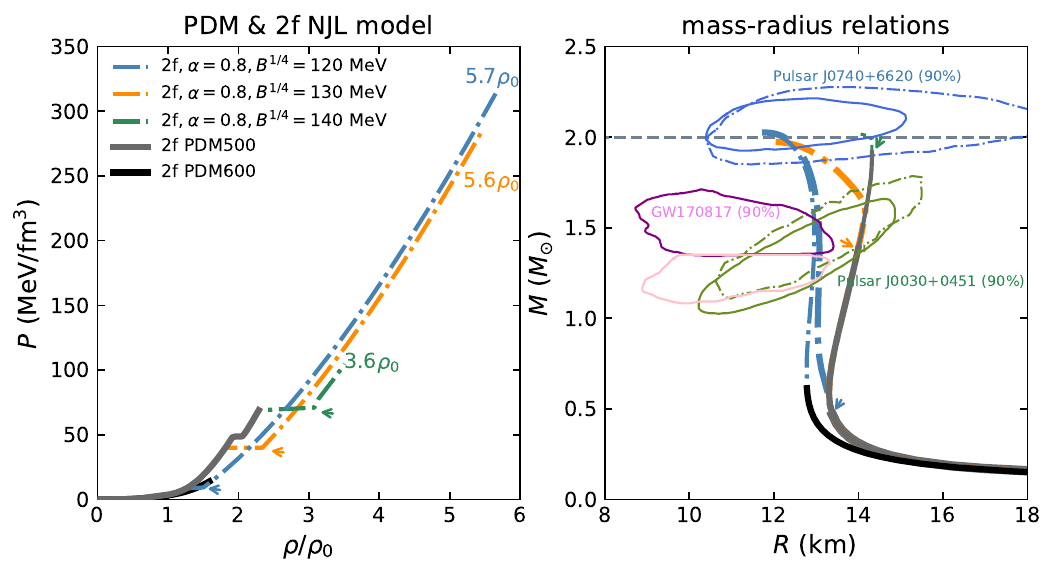}
\includegraphics[width=0.9\hsize]{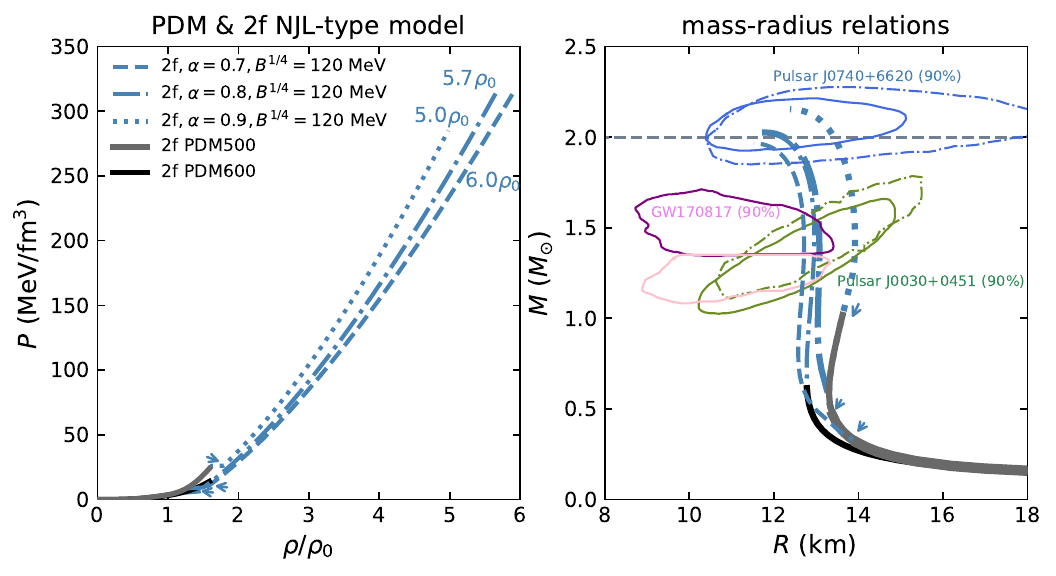}
\caption{Upper panel: The influence of the vacuum pressure $B^{1/4}$ on the results of the hybrid EOS and on the properties of hybrid stars with a two-flavor quark core. Pressure versus density (in units of nuclear saturation density $\rho_0$) for different choices of NJL parameter sets and PDM with different $m_0$ are shown in the upper left panel. The central densities $\rho_{\rm center}$ of the corresponding maximum mass hybrid stars are given. The mass-radius relations for the corresponding parameter sets are presented in the right panel. The arrows indicate the position where the quark matter begins to appear. The available mass-radius constraints from the NICER mission (PSR J0030 + 0451 ~\cite{2019ApJ...887L..24M,2019ApJ...887L..21R} and PSR J0740 + 6620 ~\cite{2021ApJ...918L..28M,2021ApJ...918L..27R}) at the 90$\%$ confidence level and the binary tidal deformability constraint from LIGO/Virgo (GW170817~\cite{2017PhRvL.119p1101A,2018PhRvL.121p1101A}) at the 90$\%$ confidence level are also shown together. Lower panel: The results of the effects of $\alpha$ are displayed with several specific parameter sets. The size of quark cores obtained are: (Upper panel) $1.51\msun$ ($\alpha=0.8, B^{1/4}=120\mev, \rm PDM500$), $0.53\msun$ ($\alpha=0.8, B^{1/4}=130\mev, \rm PDM500$), $0.08\msun$ ($\alpha=0.8, B^{1/4}=140\mev, \rm PDM600$); (Lower panel) $1.07\msun$ ($\alpha=0.7, B^{1/4}=120\mev, \rm PDM500$), $1.51\msun$ ($\alpha=0.8, B^{1/4}=120\mev, \rm PDM500$), $1.07\msun$ ($\alpha=0.9, B^{1/4}=120\mev, \rm PDM500$). The corresponding properties of the star are displayed in Table~\ref{table:PDM_2f_NJL} in detail.
}
\label{fig:eosMR}
\end{figure*}  

\begin{table*}
\centering
\caption{With PDM for hadronic phase and modified NJL model for two-flavor quark phase, the position of the deconfinement phase transition $\mu_{\rm de}$, the obtained values of the center density $\rho_{\rm center}(\rho_0)$, the corresponding deconfinement density $\rho_{\rm de}(\rho_0)$, the maximum mass $M_{\rm TOV}(M_{\odot})$ and the mass of the quark core $M_{\rm core}$ of the hybrid stars are presented for typical parameter sets.
}
                  \vskip+2mm
\renewcommand\arraystretch{1.5}
\begin{ruledtabular}
\begin{tabular*}{\hsize}{@{}@{\extracolsep{\fill}}lcccccccc@{}}
PDM500 [2f,\textbf{$\alpha=0.8$}] &  $B^{1/4}=120\mev$ & $B^{1/4}=130\mev$  & $B^{1/4}=140\mev$ & [2f, $B^{1/4}=120\mev$] & $\alpha=0.7$ & $\alpha=0.8$ &$\alpha=0.9$ &\\
\hline $\mu_{\rm de}(\rm GeV)$  & $1.00$ & $1.13$ &  $1.22$ &  $\mu_{\rm de}(\rm GeV)$  &$0.99 $ & 1.00 &1.01 &\\
\hline $\rho_{\rm center}/\rho_0$ & $5.77$ & $5.69$& $ 3.57$& $\rho_{\rm center}/\rho_0 $ & $6.01$   &5.77 &5.04 &\\
\hline  $\rho_{\rm de}/\rho_0$  & $1.26$ & $1.78 $ &  $2.29 $ &  $\rho_{\rm de}/\rho_0 $  &$1.21$&1.26 &1.59 &\\
\hline $M_{\rm TOV}/M_{\odot}$ & $2.03$ & $1.97$ & $2.02$& $M_{\rm TOV}/M_{\odot} $ & $1.97 $   &2.03 &2.15 &\\
\hline  $M_{\rm core}/M_{\odot}$ & $1.52 $ & $0.53 $ &  $0.08$ &  $M_{\rm core}/M_{\odot} $  & $1.63 $ &1.52 &1.07&\\
\hline\hline PDM600 [2f,\textbf{$\alpha=0.8$}] &  $B^{1/4}=120\mev$ & $B^{1/4}=130\mev$  & $B^{1/4}=140\mev$ & [2f, $B^{1/4}=120\mev$] & $\alpha=0.7$ & $\alpha=0.8$ &$\alpha=0.9$ &\\
\hline $\mu_{\rm de}(\rm GeV)$  & $1.02$ & $1.23$ &  $1.37 $ &  $\mu_{\rm de}(\rm GeV)$  &$0.98 $ & 1.02 &1.33 &\\
\hline $\rho_{\rm center}/\rho_0$ & $5.78$ & $5.48$& $ 4.73 $& $\rho_{\rm center}/\rho_0 $ & $5.94 $   &5.78 &5.28&\\
\hline  $\rho_{\rm de}/\rho_0$  & $1.59$ & $2.80 $ &  $3.78  $ &  $\rho_{\rm de}/\rho_0 $  &$1.24$ &1.59 &3.60&\\
\hline $M_{\rm TOV}/M_{\odot}$ & $2.02$ & $2.00$ & $2.06 $& $M_{\rm TOV}/M_{\odot} $ & $1.96 $   &2.02 &2.13&\\
\hline  $M_{\rm core}/M_{\odot}$ & $1.39 $ & $0.23 $ &  $0.02$ &  $M_{\rm core}/M_{\odot} $  & $1.63 $ &1.39 &0.13 &\\
\end{tabular*}
\end{ruledtabular}
    \vspace{-0.4cm}
\label{table:PDM_2f_NJL}
\end{table*}

\subsection{Hybrid star with large two-flavor quark matter core}\label{Sec: Hybrid star structure}
The features discussed above are further elucidated by examining the gravitational mass-radius relations under varying vacuum bag constants $B^{1/4}$ and fierz-transformed vector interactions characterized by $\alpha$. Fig.~\ref{fig:eosMR} shows the mass $M(\msun)$ of hybrid stars as a function of radius $R$ and the corresponding center densities $\rho_c$ for several selected cases. In this analysis, we focus on the parameters that influence the possible existence of large quark cores. Constraints from astronomical observations are discussed in the following sections.

\subsubsection{The influence of $B^{1/4}$}
Figure~\ref{fig:eosMR} presents the mass-radius relations for hybrid stars containing two-flavor quark cores with typical parameter sets. The upper panel emphasizes the influence of the vacuum pressure $B^{1/4}$ with relatively large $\alpha$, which corresponds to the large contribution from exchange interactions. The endpoints of the $P(\rho)$ curves correspond to the central density of the stars. The positions of deconfinement transitions are indicated with colored arrows on the mass-radius curves, providing a clear visualization of the onset of quark matter cores within the hybrid star configurations.  

As discussed in Sec.~\ref{Sec: Hybrid star EOS}, a vacuum pressure greater than $B^{1/4} = 115~\mathrm{MeV}$ is necessary to ensure an intersection between the hadronic and quark matter EOSs. For $\alpha = 0.8$ at $B^{1/4} = 120~\mathrm{MeV}$, constructed with nuclear EOS for PDM500, the deconfinement chemical potential is found to be $\mu_{\rm de} = 1.0~\mathrm{GeV}$. At this transition point, the energy density discontinuity between the hadronic and quark phases is relatively small, allowing the hybrid star to reach a maximum mass of $M_{\rm TOV} = 2.03~M_\odot$. Notably, the corresponding two-flavor quark matter core is large, which has a mass of $M_{\rm core} = 1.51~M_\odot$. Previous studies also found there exist quark core~\cite{1999PhRvC..60b5801S,2020PhRvD.101l3030F,2024PhRvC.110d5802G} but such a huge one is rare. 
Increasing the vacuum pressure to $B^{1/4} = 130~\mathrm{MeV}$ results in a larger energy density discontinuity between these two phases. This leads to a smaller hybrid star with a reduced maximum mass of $M_{\rm TOV} = 1.97~M_\odot$ and a correspondingly smaller quark core $\sim 0.53\msun$. The position of the deconfinement chemical potential $\mu_{\rm de}$ is highly sensitive to changes in $B^{1/4}$, highlighting the direct influence of the vacuum pressure on the phase transition properties. Further increasing the vacuum pressure to $B^{1/4} = 140~\mathrm{MeV}$ shifts the deconfinement transition to even higher chemical potentials. At this value, the hybrid star achieves a maximum mass of \(M_{\rm TOV} = 2.02~M_\odot\), while the quark core becomes significantly smaller, with a mass of only $M_{\rm core} = 0.08~M_\odot$, which is demonstrated in Table~\ref{table:PDM_2f_NJL}. A comparison between the results for $B^{1/4} = 130~\mathrm{MeV}$ and $B^{1/4} = 140~\mathrm{MeV}$ suggests that when the phase transition occurs at very high densities, the star is able to sustain more mass in the hadronic phase before transitioning to the softer quark matter phase, resulting in a higher maximum mass but smaller quark cores. These findings indicate that, as the deconfinement transition shifts to higher baryon densities, the hadronic phase becomes increasingly dominant in the internal structure of the hybrid star. 
Interestingly, the impact of vacuum pressure on the stiffness of the hybrid star EOS is complex: both very large and very small quark cores can support high maximum masses. This intricate behavior underscores the crucial role of the vacuum pressure $B^{1/4}$ in shaping the mass-radius relations and internal composition of hybrid stars.

Comparing the cases of $\rm PDM 500$ and $\rm PDM 600$, while remaining the quark matter EOS unchanged with $\alpha=0.8$ and $B^{1/4}=120\mev$, it becomes clear that the hadronic matter EOS predominantly determines the radius of the hybrid star in the low-density regime. Reducing the invariant mass $m_0$ stiffens the hadronic matter EOS, leading to a slightly stiffer hybrid star EOS. This increased stiffness allows the star to support slightly higher maximum masses from $2.02\msun$ to $2.03\msun$, as the delayed deconfinement phase transition expands the region dominated by hadronic matter. 

\subsubsection{The impact of $\alpha$}
The influence of the parameter $\alpha$ on the phase transition point and hybrid star configuration is depicted in the lower panel of Fig.~\ref{fig:eosMR}. The parameter $\alpha$ controls the weight of Fierz-transformed interaction channels in the modified NJL model. An increase in $\alpha$ enhances the contribution of vector interactions, resulting in a stiffer quark matter EOS. This stiffening pushes the deconfinement transition chemical potential $\mu_{\rm de}$ to higher baryon number densities, allowing the neutron star to sustain a more massive hadronic matter shell before transitioning to the quark matter phase. 

In particular, for the fixed vacuum pressure of $B^{1/4} = 120~\mathrm{MeV}$, the effects of $\alpha$ on the hybrid star properties become particularly evident. Employing the PDM500 constructed with two-flavor quark matter EOS at $\alpha = 0.7$, the deconfinement transition occurs at a relatively low baryon chemical potential, $\mu_{\rm de} = 0.99~\mathrm{GeV}$, resulting in a hybrid star with a maximum mass of $M_{\rm TOV} = 1.97~M_\odot$ and a substantial two-flavor quark core of $M_{\rm core} = 1.63~M_\odot$. Increasing $\alpha$ to $0.8$ stiffens the quark matter EOS, raising the deconfinement transition chemical potential to higher values. In this case, the hybrid star reaches a maximum mass of $M_{\rm TOV} = 2.03~M_\odot$, with a slightly smaller two-flavor quark core of $M_{\rm core} = 1.52~M_\odot$. A further increase to $\alpha = 0.9$ produces an even stiffer quark matter EOS, which supports a larger maximum mass of $M_{\rm TOV} = 2.15~M_\odot$. The increased stiffness of quark matter EOS delayed the position of $\mu_{\rm de}$, reducing the size of the quark core with $M_{\rm core} = 1.07~M_\odot$. 

These results highlight the sensitivity of hybrid star properties to the parameter $\alpha$, which governs the interplay between scalar and vector interaction channels in the modified NJL model. Notably, except for the case of $\alpha = 0.9$, the configurations suggest that GW170817 could potentially be a hybrid star containing a large quark core. The implications of this observation are significant, as it reinforces the possibility of quark matter existing in the cores of neutron stars observed in gravitational wave events GW170817.

\subsection{Hybrid star with (2+1)-flavor quark core}\label{section: 3flavor quark core}
\begin{figure*}
\centering
{\includegraphics[width=0.9\textwidth]{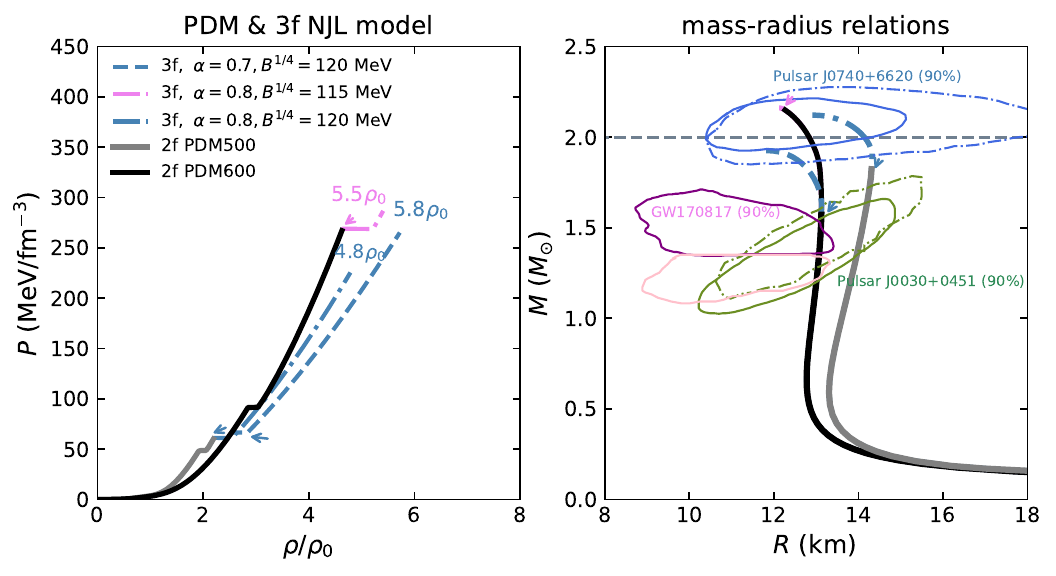}}
\caption{ Pressure versus density (left panel) in units of nuclear saturation density $\rho_0$ and mass-radius relations (right panel) with strange quark matter core for selected parameter sets. The arrows in the right panel indicate the position where the quark matter begins to appear. The size of quark cores obtained are: $0.35\msun$ ($\alpha=0.7, B^{1/4}=120\mev, \rm PDM600$), $0.01\msun$ ($\alpha=0.8, B^{1/4}=115\mev, \rm PDM600$), $0.28\msun$ ($\alpha=0.8, B^{1/4}=120\mev, \rm PDM500$). The center densities corresponding to the maximum mass are: $5.8\rho_0$ ($\alpha=0.7, B^{1/4}=120\mev, \rm PDM600$), $5.5\rho_0$ ($\alpha=0.8, B^{1/4}=115\mev, \rm PDM600$), $4.8\rho_0$ ($\alpha=0.8, B^{1/4}=120\mev, \rm PDM500$). The available mass-radius constraints from the NICER mission (PSR J0030 + 0451 ~\cite{2019ApJ...887L..24M,2019ApJ...887L..21R} and PSR J0740 + 6620 ~\cite{2021ApJ...918L..28M,2021ApJ...918L..27R}) at the 90$\%$ confidence level and the binary tidal deformability constraint from LIGO/Virgo (GW170817~\cite{2017PhRvL.119p1101A,2018PhRvL.121p1101A}) at the 90$\%$ confidence level are also shown together.
}
\label{fig:eos_MR3f}
\end{figure*}

If the hybrid EOS is constructed using the modified (2+1)-flavor NJL model for the quark phase, it becomes challenging for hybrid stars to host a substantial strange quark core larger than $\sim 1.0~M_\odot$, despite the existence of parameter space for stable hybrid stars with strange quark cores. This difficulty arises from the somewhat complex behavior of the intersections between the two phases in the $P(\mu_B)$ curves, as discussed in Sec.~\ref{Sec: Hybrid star EOS}. In this case, It is important to increase the chiral invariant mass from $500\mev$ to $600\mev$, which results in a shrinking radius of the hybrid star at the intermediate density region to satisfy the constraint from the GW170817 event. Specifically, for the PDM model with $m_0 = 600~\mathrm{MeV}$, achieving a single intersection of the $P(\mu_B)$ curves requires a vacuum pressure $B^{1/4}$ greater than $113~\mathrm{MeV}$ for $\alpha = 0.8$. As shown in Fig.~\ref{fig:eos_MR3f}, at $B^{1/4} = 115~\mathrm{MeV}$, this leads to a large deconfinement chemical potential, $\mu_{\rm de} = 1.52~\mathrm{GeV}$, and a correspondingly small strange quark core mass of $M_{\rm core} = 0.01~M_\odot$. Increasing the vacuum pressure further significantly raises the value of $\mu_{\rm de}$, making it increasingly difficult for hybrid stars to host a substantial strange quark core. On the other hand, reducing the weight of the exchange-channel interactions, $\alpha$, softens the strange quark matter EOS, which reduces the hybrid star's ability to support massive stars of $\sim 2.0~M_\odot$. Thus, in this scenario, it is inherently difficult for a hybrid star to host a large strange quark core. Moreover, the radius of a $1.4~M_\odot$ hybrid star decreases by approximately $1~\mathrm{km}$ when $m_0$ is increased from $500~\mathrm{MeV}$ to $600~\mathrm{MeV}$. This adjustment brings the mass-radius relation into better agreement with observational constraints from both LIGO/Virgo and NICER.

%

\begin{table*}
\centering
\caption{With PDM for hadronic phase and modified (2+1)-flavor NJL model for quark phase, the position of the deconfinement phase transition $\mu_{\rm de}$, the center density $\rho_{\rm center}(\rho_0)$ of the star, the corresponding deconfinement density $\rho_{\rm de}(\rho_0)$, the maximum mass $M_{\rm TOV}(M_{\odot})$ and the mass of the quark core $M_{\rm core}$ of the hybrid stars for typical parameter sets. 
}
                  \vskip+2mm
\renewcommand\arraystretch{1.5}
\begin{ruledtabular}
\begin{tabular*}{\hsize}{@{}@{\extracolsep{\fill}}lcccccccc@{}}
PDM500 [3f,\textbf{$\alpha=0.8$}] & $B^{1/4}=115\mev$ & $B^{1/4}=120\mev$ & $B^{1/4}=130\mev$  &  [3f, $B^{1/4}=120\mev$] & $\alpha=0.7$ & $\alpha=0.8$ &$\alpha=0.9$ &\\
\hline $\mu_{\rm de}(\rm GeV)$  &$1.10 $ & $1.19$ & $1.30$ &    $\mu_{\rm de}(\rm GeV)$  &$1.06 $ & 1.19 &/ &\\
\hline $\rho_{\rm center}/\rho_0$  & $5.15 $& $4.79$ & $3.75$&  $\rho_{\rm center}/\rho_0 $ & $6.11 $   &4.79 &/ &\\
\hline  $\rho_{\rm de}/\rho_0$  & $1.681 $ & $2.19$ & $2.62 $ &   $\rho_{\rm de}/\rho_0 $  &$1.54 $&2.19 &/&\\
\hline $M_{\rm TOV}/M_{\odot}$ & $2.10  $& $2.12$ & $2.25$ &  $M_{\rm TOV}/M_{\odot} $ & $1.90  $   &2.12 &/ &\\
\hline  $M_{\rm core}/M_{\odot}$ & $0.83  $ &  $0.28 $ & $0.04 $ &   $M_{\rm core}/M_{\odot} $  & $0.90  $ &0.28 &/ &\\
\hline\hline PDM600 [3f,\textbf{$\alpha=0.8$}] & $B^{1/4}=115\mev$ & $B^{1/4}=120\mev$ & $B^{1/4}=130\mev$  &  [3f, $B^{1/4}=120\mev$] & $\alpha=0.7$ & $\alpha=0.8$ &$\alpha=0.9$ &\\
\hline $\mu_{\rm de}(\rm GeV)$  & $1.52$ & $1.57 $ & $\ 1.64$ &    $\mu_{\rm de}(\rm GeV)$  &$1.18  $ & 1.57 &/ &\\
\hline $\rho_{\rm center}/\rho_0$ & $ 5.46 $& $5.73 $ & $3.75 $&  $\rho_{\rm center}/\rho_0 $ & $5.79 $   &5.73 &/ &\\
\hline  $\rho_{\rm de}/\rho_0$  & $4.63 $ &  $4.86$ & $5.23 $ &   $\rho_{\rm de}/\rho_0 $  &$2.52 $&4.86 &/&\\
\hline $M_{\rm TOV}/M_{\odot}$ & $2.16 $& $2.17 $ & $2.19 $ &  $M_{\rm TOV}/M_{\odot} $ & $1.93 $   &2.17 &/ &\\
\hline  $M_{\rm core}/M_{\odot}$ & $0.01 $ & $0.002 $ & $0.001 $ &    $M_{\rm core}/M_{\odot} $  & $0.35  $ &0.002 &/ &\\
\end{tabular*}
\end{ruledtabular}
    \vspace{-0.4cm}
\label{table:PDM_3f}
\end{table*}

\begin{figure*}
\centering
{\includegraphics[width=0.48\textwidth]{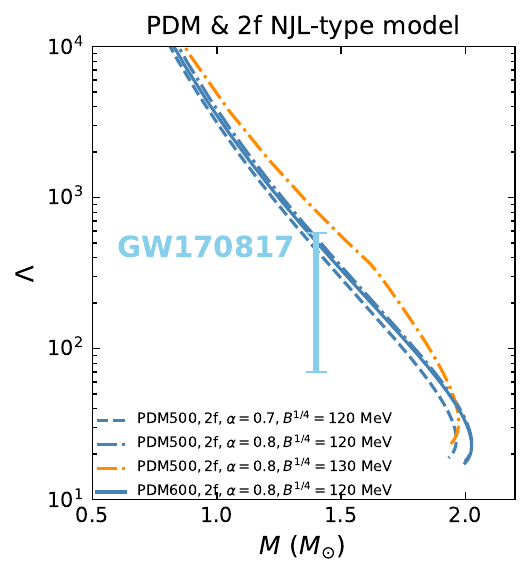}}
{\includegraphics[width=0.48\textwidth]{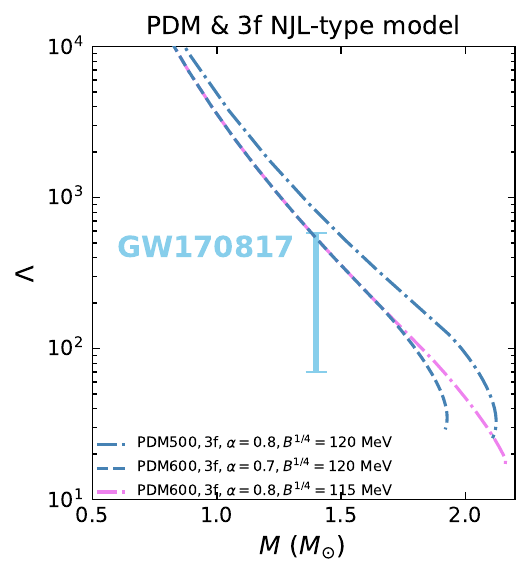}} 
\caption{The dimensionless tidal deformability \(\Lambda\) as functions of stellar mass \(M\, (M_{\odot})\) are shown for hybrid stars with two- and three-flavor quark cores. The parameter sets used are the same as those presented in Figs.~\ref{fig:eosMR} and \ref{fig:eos_MR3f} of this paper. The constraint from the GW170817 event places \(\Lambda\) in the range \(70 < \Lambda < 580\)~\cite{2017PhRvL.119p1101A,2018PhRvL.121p1101A}, which is illustrated as a blue error bar.}
\label{fig:tidal} 
\end{figure*}

\subsection{Tidal deformability}
The binary neutron star merger event GW170817 provides a constraint on dimensionless tidal deformability \(\Lambda\) for a \(1.4\,M_{\odot}\) compact star, which directly place $\Lambda$ in the range \(70 < \Lambda < 580\)~\cite{2017PhRvL.119p1101A,2018PhRvL.121p1101A}. In the following, we directly compute \(\Lambda\) for hybrid stars containing both nonstrange and strange quark cores, using geometric units with $G =c= 1$. To linear order, the tidal deformability $\lambda$, which characterizes the response of a neutron star to an external tidal field, is defined as the ratio of the induced mass quadrupole moment $Q_{i j}$ to the applied tidal field $\mathcal{E}_{i j}$, $Q_{i j}=-\lambda \mathcal{E}_{i j}$, and is related to the $l=2$ dimensionless tidal Love number $k_2$, expressed as
\begin{equation}
\begin{aligned}
\lambda &=\frac{2}{3} k_2 R^5\ , \\
\Lambda &= \lambda/M^5\ .
\end{aligned}
\end{equation}
Here, $M$ and $R$ are the mass and radius of the star, respectively, and $\Lambda$ is the dimensionless tidal deformability.

Following the method developed in Refs.~\cite{2008ApJ...677.1216H,2009PhRvD..80h4035D,2010PhRvD..82b4016P,2020PhRvD.102b8501T}, we should solve the following TOV equations 
\begin{equation}
\begin{aligned}
\frac{\mathrm{d} P(r)}{\mathrm{d} r}&=-\frac{(\epsilon+P)\left(M+4 \pi r^3 P\right)}{r(r-2 M)}\ , \\
\frac{\mathrm{d} M(r)}{\mathrm{d} r}&=4 \pi r^2 \epsilon\ , \\
\end{aligned}
\end{equation}
and
\begin{widetext}
\begin{equation}
\begin{aligned}
\frac{\mathrm{d} H(r)}{\mathrm{d} r}&=\beta\ ,\\
\frac{d\beta(r)}{dr} &=\  2 \left(1 - \frac{2M}{r} \right)^{-1} H \bigg[
    -2\pi \left( 5\epsilon + 9P + f(\epsilon + P) \right)
    + \frac{3}{r^2} + 2 \left(1 - \frac{2M}{r} \right)^{-1} \left( \frac{M}{r^2} + 4\pi r P \right)^2\bigg] \\
&+ \frac{2\beta}{r} \left(1 - \frac{2M}{r} \right)^{-1} \left(
    -1 + \frac{M}{r} + 2\pi r^2 (\epsilon - P)
\right)\ ,\label{eq:tidal_Defor}
\end{aligned}
\end{equation}
\end{widetext}
simultaneously to calculate the tidal deformability with the EOSs. Here, for slow changes in matter configurations, $f$ is given by
\begin{equation}
f = \frac{d\epsilon}{dP}\ . 
\end{equation} 

For the internal solution, the $l=2$ tidal Love number $k_2$ is expressed as:
\begin{equation}
\begin{aligned}
k_2& = \frac{8C^5}{5}(1 - 2C)^2 \big[ 2 + 2C(y_R - 1) - y_R \big]\\
&\times \Big\{ 
2C \big[6 - 3y_R + 3C(5y_R - 8) \big] \\
& + 4C^3 \big[13 - 11y_R + C(3y_R - 2) + 2C^2(1 + y_R) \big] \\
& + 3(1 - 2C)^2 \big[2 - y_R + 2C(y_R - 1) \big] \ln(1 - 2C)
\Big\}^{-1},
\end{aligned}
\end{equation}
in which $C=M/R$ is the compactness of the star, and $y_R =  r\beta(r)/H(r)|_{r=R}$. 

In the presence of a finite energy density discontinuity, there is a jump of $\Delta \epsilon$ in the energy density at constant pressure $P_{\rm tr}$. Hence $f=\mathrm{d}\epsilon/{\mathrm{d} P}$ displays a delta-function behavior across the point of discontinuity, which is expressed as
\begin{equation}
f=\left.\frac{\mathrm{d} \epsilon}{\mathrm{~d} P}\right|_{P \neq P_{\mathrm{tr}}}+\delta\left(P-P_{\mathrm{tr}}\right) \Delta \epsilon\ .
\end{equation}
This leads to an extra term for the solution of $y(r)$ across $r_{\mathrm{tr}}$ as discussed in Refs.~\cite{2009PhRvD..80h4035D,2010PhRvD..82b4016P,2019PhRvD..99h3014H,2020PhRvD.102b8501T,2021PhRvD.104d3002L}, which is given by
\begin{equation}
y\left(r_{\mathrm{tr}}^{+}\right)-y\left(r_{\mathrm{tr}}^{-}\right)=-\frac{4 \pi r_{\mathrm{tr}}^3 \Delta \epsilon}{M\left(r_{\mathrm{tr}}\right)+4 \pi r_{\mathrm{tr}}^3 P\left(r_{\mathrm{tr}}\right)}\ .
\end{equation}
Here, $\Delta \epsilon=\epsilon\left(r_{\text {tr }}^{-}\right)-\epsilon\left(r_{\text {tr }}^{+}\right)$, and $r_{\text {tr }}^{ \pm}=r_{\mathrm{tr}} \pm \delta r$, in which $r_{\mathrm{tr}}$ represents the position where this first-order hadron-quark phase transition happens, and $\delta r$ is an infinitesimal distance around $r_{\mathrm{tr}}$. The values of $\Lambda$ were calculated using the code developed by Andrea Maselli~\cite{A. Masellicode}, which has also been used to compute $\Lambda$ for cases involving sharp first-order phase transitions, as discussed in Ref.~\cite{2023PhRvD.107f3034B}.

In Fig.~\ref{fig:tidal}, we compare our calculated values of $\Lambda$ at \(1.4\,M_{\odot}\), obtained from the PDM and NJL models across several parameter sets, with the GW170817 observational constraint.  Our results show that the mass–radius relations derived from models consistent with the LIGO/Virgo constraint at the $90\%$ confidence level also yield tidal deformabilities within the allowed range. For a given quark matter EOS, increasing $m_0$ leads to a reduction in $\Lambda$. Therefore, the tidal deformability constraint from GW170817 favors larger values of $m_0=600\mev$, corresponding to a slightly softer hadronic EOS. Compared with the discontinuous behavior of $\Lambda$ reported in Ref.~\cite{2019PhRvD..99h3014H}, where the authors employ the SFHo and Dirac–Brueckner–Hartree–Fock (DBHF) EOS for low-density matter and adopt the constant-sound-speed (CSS) model for high-density quark matter to construct hybrid star configurations with sharp first-order phase transitions, our results are different. In their work, it was found that if the resulting hybrid EOS gives rise to twin-star or triplet-star solutions with an intermediate unstable branch, clear discontinuities can appear in the tidal deformability versus gravitational mass ($\Lambda$–$M$) relation due to the existence of the intermediate unstable branch. In contrast, in our study based on the PDM and modified NJL models, we do not observe twin-star configurations, and the tidal deformabilities we compute remain continuous, which has also been found in the study of hybrid stars with sharp phase transitions in Ref.~\cite{2023PhRvD.107f3034B}.

\subsection{Stable parameter space with quark core}
\begin{figure*}
\centering
{\includegraphics[width=0.49\textwidth]{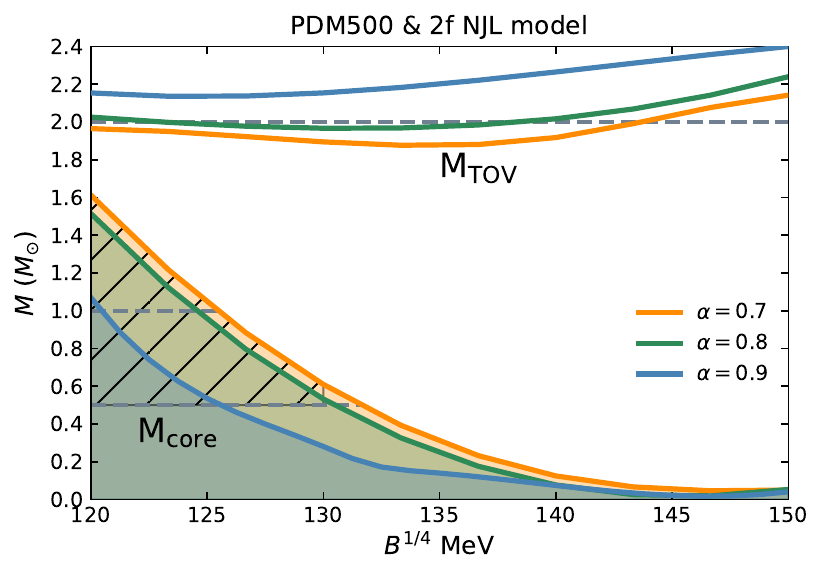}}
{\includegraphics[width=0.49\textwidth]{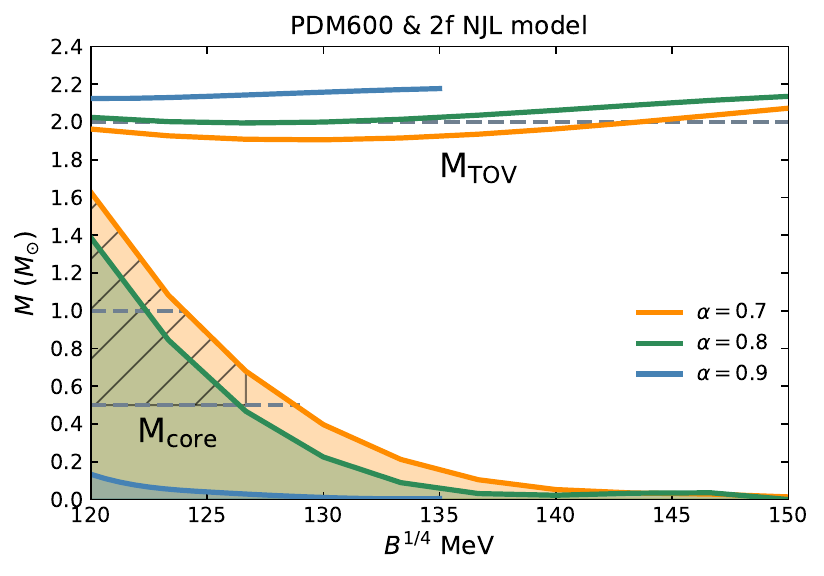}}
{\includegraphics[width=0.49\textwidth]{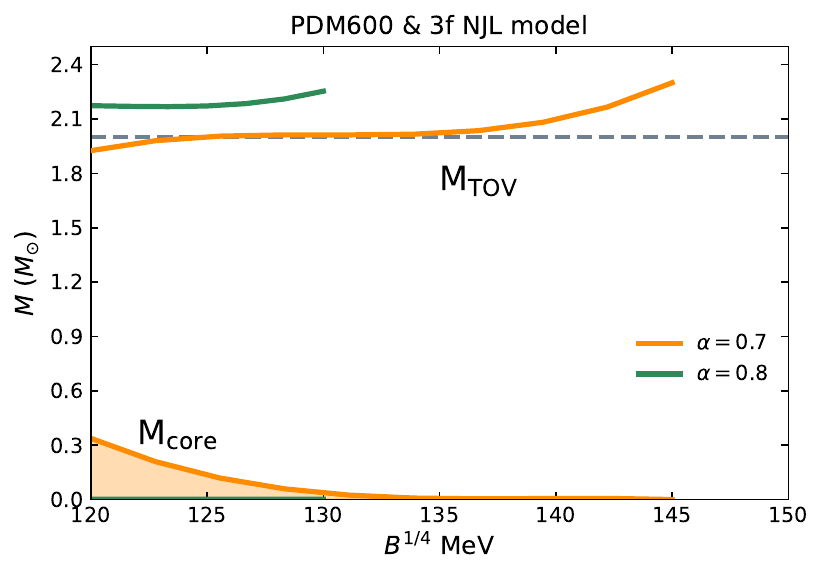}}
\caption{
Upper panel: The maximum mass of hybrid stars and the corresponding masses of nonstrange quark cores as functions of vacuum pressure $B^{1/4}$, using the two-flavor modified NJL model for the quark phase and the PDM model for the hadronic phase with $m_0 = 500~\mathrm{MeV}$ and $m_0 = 600~\mathrm{MeV}$, respectively. Different colored curves represent various cases of NJL parameters for $\alpha$. Lower panel: The corresponding results for strange quark matter cores in hybrid stars.
}
\label{fig:PDM500_2fquarkcore}
\end{figure*}

In this section, we further present the maximum mass and the mass of the quark core as functions of the NJL model parameter, as shown in Fig.~\ref{fig:PDM500_2fquarkcore}. The analysis is performed for two cases of the PDM model with different chiral invariant masses: $m_0 = 500~\mathrm{MeV}$ and $m_0 = 600~\mathrm{MeV}$.

Specifically, we highlight the maximum mass and the masses of both nonstrange and strange quark cores as functions of vacuum pressure for various NJL model parameters at different values of $\alpha$. As shown in Fig.~\ref{fig:PDM500_2fquarkcore}, the colored shaded regions correspond to the size of quark cores as functions of vacuum pressure for a given value of $\alpha$. We find that the two-flavor quark matter occupies a larger parameter space for appearing in the interior of neutron stars compared to strange quark matter. For hybrid stars with nonstrange quark matter cores, $\alpha$ must exceed $0.7$ to ensure a sufficiently stiff hybrid EOS to support the observations of massive mass $\sim 2\msun$. At $\alpha = 0.8$, a broad range of vacuum pressures from $B^{1/4} = 120~\mathrm{MeV}$ to $125~\mathrm{MeV}$ allows the nonstrange quark core to exceed $\sim 1.0~M_\odot$. However, for $\alpha = 0.9$, the range of $B^{1/4}$ is significantly narrower due to the high transition densities. In comparison, when adopting the EOS of softer hadronic matter described by PDM 600 with the same two-flavor NJL model parameters, the size of the nonstrange quark core is reduced overall. This highlights the sensitivity of hybrid star properties to both the vacuum pressure and the stiffness of the hadronic EOS.

For a hybrid star with strange quark core,  as we mentioned in Sec.~\ref{section: 3flavor quark core}, a relatively large $m_0$ should be employed to ensure the radius of a star can reconcile with the radius reported from the gravitational wave observation of the binary neutron star merger event GW170817. Therefore, we display the maximum mass and quark cores employing the PDM600 for nuclear matter and (2+1)-flavor NJL model for quark matter at high densities. As shown in the lower panel in Fig.~\ref{fig:PDM500_2fquarkcore}, there exist a narrow parameter space to allow the hybrid star to host a quark core, and no parameter space that supports the hybrid star to have a large strange quark core beyond $\sim 1\msun$.

From the discussion above, the results indicate that the quark matter EOS must be sufficiently stiff to allow the formation of a quark core within a neutron star. Generally speaking, increasing $\alpha$ enhances the vector interactions in the exchange channels, leading to a stiffer quark matter EOS, which results in a larger maximum mass for the hybrid star but a smaller quark core. For a fixed $\alpha$, increasing the vacuum pressure $B^{1/4}$ initially causes a slight decrease in the maximum mass of the hybrid star. However, as $B^{1/4}$ increases further, the maximum mass begins to rise due to the competition between the hadronic phase and the quark phase. If the bag constant becomes too large, the quark matter EOS softens excessively, reducing the pressure support from the quark core, which becomes insufficient to counteract gravitational collapse, ultimately leading to an unstable configuration.

\section{Conclusions and summary}\label{sec:summary}
The phase state at supranuclear densities has been a challenging topic that is directly related to neutron star physics. Due to the possible appearance of various exotic degrees of freedom, such as hyperons, kaons, Delta isobars, or deconfined quarks, the composition of the compact stars is complicated. In this work, we systematically investigate hybrid star configurations with two- or three-flavor quark matter cores employing theoretical models that both respect the chiral symmetry of QCD, i.e., a PDM-derived hadronic EOS and a modified NJL model EOS. The NJL model Lagrangian incorporates both original interaction channels and their Fierz-transformed counterparts. The parameter $\alpha$, which ranges from 0 to 1, is treated as a free variable to be constrained by astronomical observations due to the lack of experimental data. Instead of exploring the crossover transition scenario within the framework of PDM and NJL-type models~\cite{2021PhRvC.103d5205M,2021PhRvC.104f5201M,2022PhRvC.106f5205G,2023Symm...15..745M,2024PhRvC.109f5807G,2024arXiv240318214G}, our study uses the assumption of a sharp first-order phase transition employing a Maxwell construction.
 
We find that the chiral invariant mass $m_0$ in the PDM plays a crucial role in determining the stiffness of the hadronic EOS, while the parameters $\alpha$ and $B$ in the NJL model govern the stiffness of the quark matter EOS. The interplay between the non-perturbative interactions within the PDM and NJL model controls the location of the phase transition. In the three-flavor NJL model, the inclusion of strange quarks introduces complexity in the pressure vs. baryon chemical potential relation, resulting in multiple intersections between the hadronic and strange quark matter EOSs.  

For hybrid stars with two-flavor quark matter cores, our results reveal substantial parameter space supporting large quark cores exceeding $\sim 1 M_\odot$ for PDM invariant masses of both $500 \, \mathrm{MeV}$ and $600 \, \mathrm{MeV}$. These stellar configurations satisfy the mass and radius constraints from observations of pulsars J0740+6620 and J0030+0451, as well as gravitational wave event GW170817. The result obtained here is different from Refs.~\cite{1999PhRvC..60b5801S,2024PhRvC.110d5802G,2025ApJ...980..231L}, in which the typical neutron stars with masses around $1.4\msun$ do not possess any deconfined quark matter in their center. In particular, Ref.~\cite{2025ApJ...980..231L} studied the hybrid stars with nonstrange quark cores, employing NJL-type model for the quark phase and APR for the hadronic phase, and found out that the nonstrange quark cores can possess $0.026-0.04\msun$. The discrepancy can be generated from the chiral symmetry properties within PDM. For hadronic matter with $m_0 = 500 \, \mathrm{MeV}$, the maximum mass reconcilable with current observations is $2.03 \msun$, with a two-flavor quark core mass of $1.52 \msun$ at $B^{1/4} = 120 \, \mathrm{MeV}$ and $\alpha = 0.8$. Higher values of $\alpha$ or $B^{1/4}$ shift the phase transition to higher chemical potentials, resulting in radii inconsistent with constraints from GW170817. For $m_0 = 600 \, \mathrm{MeV}$, the maximum mass increases to $2.18 \msun$, with a much smaller quark core mass of $0.005 M_\odot$ at $B^{1/4} = 135 \, \mathrm{MeV}$ and $\alpha = 0.9$. 

In contrast, for hybrid stars with three-flavor quark matter cores, a larger invariant mass of $m_0 \approx 600 \, \mathrm{MeV}$ is necessary to satisfy the tidal deformability constraint for a $1.4 \msun$ star from GW170817 event. This requirement suggests that the intermediate-density EOS should be moderately soft. To remain consistent with maximum mass constraints, the quark matter EOS must be sufficiently stiff, necessitating a relatively high value of $\alpha$ in the modified NJL model to enhance vector interactions. The vacuum bag constant $B$ plays a pivotal role in determining the phase transition chemical potential $\mu_{\rm de}$. Increasing $B$ leads to higher maximum masses but smaller and less stable quark cores.
By comparing our theoretical results with observational constraints from gravitational wave measurements (LIGO/Virgo) and pulsar observations (NICER), we find that the maximum mass for a hybrid star with a strange quark core reaches approximately $2.2 M_\odot$ for $B^{1/4} = 125 \, \mathrm{MeV}$ and $\alpha = 0.85$ at $m_0 = 600 \, \mathrm{MeV}$. In contrast, for $m_0 = 500 \, \mathrm{MeV}$, hybrid star configurations struggle to meet the constraints imposed by GW170817. The hybrid stars' maximum masses are found to be approximately $2.2\msun$ for both two- or three-flavor quark matter in their centers. This is also found in Refs.~\cite{2023SciBu..68..913H,2024PhRvD.109d3052F}, where they use a data-driven approach to infer that the maximum mass for a neutron star with a quark core is $2.25^{+0.08}_{-0.07}\msun$. 
Our present study serves as an essential step in validating the hybrid star model before extending it to finite-temperature simulations for supernova explosions and neutron star mergers. 

Large quark cores in hybrid stars have also been reported in studies such as Refs.~\cite{2022PhRvC.105d5808C,2023PhRvD.108k4028G,2025arXiv250100115A}, which typically adopt RMF models for the hadronic phase or employ simplified quark matter EOSs for computational convenience. In contrast, our work incorporates effective models—PDM and NJL—that both respect chiral symmetry. Given the considerable uncertainties regarding the properties of the QCD phase transition, terrestrial experiments, while effective at constraining the EOS near the nuclear saturation density, cannot definitively rule out the possibility of a phase transition occurring at densities slightly above \( n_0 \). Within our phenomenological framework, we find that a low transition density is indeed plausible, particularly for hybrid stars with nonstrange quark cores. This is because two-flavor quark matter generally exhibits a stiffer EOS than its three-flavor counterpart. As a result, for a given hadronic EOS, the stiffer nonstrange quark matter leads to the phase transition at lower densities. Looking ahead, if future observations confirm the existence of subsolar-mass compact stars, our results may indicate that such objects are more likely to be hybrid stars with nonstrange quark cores, rather than stars composed of strange quark matter.

Finally, we acknowledge several limitations of our approach. While the PDM framework offers valuable insights, it is not without challenges, such as its fundamental assumption that the $N(939)$ and $N(1535)$ baryons act as chiral partners. This identification remains phenomenological and is not conclusively supported by experimental evidence. The significant mass difference between these states necessitates the introduction of intricate symmetry breaking mechanisms, which may introduce additional model dependencies and influence our results. Moreover, extrapolating the PDM to high densities involves inherent uncertainties. As with many hadronic models, the PDM is primarily calibrated near nuclear saturation density \( \rho_0 \), due to limited knowledge of dense matter at higher densities. Consequently, its predictions at \(2\)–\(5\,\rho_0\) become increasingly speculative. In such regimes—typical of neutron star cores—hyperons are expected to appear at densities around \(2\)–\(3\,\rho_0\), as the elevated baryon chemical potential makes their formation energetically favorable~\cite{2013PhRvC..87a5804D}. The presence of hyperons could significantly alter the equation of state at high densities. Nevertheless, our present work focuses on exploring the interplay between the quark-hadron phase transition and chiral symmetry restoration, for which the nucleonic PDM provides a suitable starting point. Future studies may extend this framework to include hyperons, allowing for a more comprehensive investigation of their role in both chiral symmetry restoration and deconfinement.

\medskip
\acknowledgments
We thank Masayasu Harada, Yong-Liang Ma, and Chen Zhang for the helpful discussions. We also thank Professor David Blaschke's comments. This work is supported by the National SKA Program of China (2020SKA0120100), the National Natural Science Foundation of China (Nos. 12003047, 12133003), the Strategic Priority Research Program of the Chinese Academy of Sciences (No. XDB0550300), and the Special Funds of the National Natural Science Foundation of China (Grant No. 12447171). Bikai Gao is supported in part by JSPS KAKENHI Grant Nos.~20K03927, 23H05439, 24K07045, and JST SPRING, Grant No. JPMJSP2125.  Bikai Gao would like to take this opportunity to thank the “Interdisciplinary Frontier Next-Generation Researcher Program of the Tokai Higher Education and Research System.”

\end{document}